\documentclass[10pt]{article}
\newtheorem{thm}{Theorem}
\newtheorem{pro}{Proposition}[section]
\newtheorem{lem}{Lemma}[section]

\newtheorem{rmk}{Remark }

\catcode`@=11 \@addtoreset{equation}{section} \catcode`@=12
\usepackage{subfigure}
\usepackage[table]{xcolor}
\usepackage{graphicx}
\usepackage{amssymb,epsfig,amsmath,times,cite,bbm,lineno,color}
\usepackage{psfrag}
\usepackage{arydshln}
\title{\bf Global stability of epidemic models with imperfect vaccination and quarantine on scale-free networks}
\author{Shanshan Chen$^{1,2,3}$,  Michael Small$^{3,4}$, Xinchu Fu$^{1,}$\thanks{Corresponding author. Tel: +86-21-66132664; Fax: +86-21-66133292; Email address: xcfu@shu.edu.cn}
\\
{\it \small $^1$Department of Mathematics, Shanghai University, 99 Shangda Road, Shanghai 200444, China}\\
{\it \small $^2$Department of Computer Science, School of Electronic and Electrical Engineering,}\\
{\it \small Shanghai University of Engineering Science, 333 Longteng Road, Shanghai, China}\\
{\it \small $^3$School of Mathematics and Statistics, University of Western Australia, Crawley, 6009, Australia}\\
{\it \small $^4$Mineral Resources, CSIRO, Kensington, 6151, Australia} }

\date{}
\begin{document}

\maketitle

\begin{abstract}
\noindent Public health services are constantly searching for new ways to reduce the spread of infectious diseases, such as public vaccination of asymptomatic individuals, quarantine (isolation) and treatment of symptomatic individuals. Epidemic models have a long history of assisting in public health planning and policy making. In this paper, we introduce epidemic models including variable population size, degree-related imperfect vaccination and quarantine on scale-free networks. More specifically, the models are formulated both on the population with and without permanent natural immunity to infection, which corresponds respectively to the
susceptible-vaccinated-infected-quarantined-recovered (SVIQR) model and the susceptible-vaccinated-infected-quarantined (SVIQS) model. We develop different mathematical methods and techniques to study the dynamics of two models,
including the basic reproduction number, the global stability of disease-free and endemic equilibria.
For the SVIQR model, we show that the system exhibits a forward bifurcation.
Meanwhile, the disease-free and unique endemic equilibria are shown to be globally asymptotically stable by constructing suitable Lyapunov functions.
For the SVIQS model, conditions ensuring the occurrence of multiple endemic equilibria are derived.
Under certain conditions, this system cannot undergo a backward bifurcation.
The global asymptotical stability of disease-free equilibrium, and the persistence of the disease are proved.
The endemic equilibrium is shown to be globally attractive by using monotone iterative technique.
Finally, stochastic network simulations yield quantitative agreement with the
deterministic mean-field approach.

\vspace{0.2cm}

\end{abstract}


\section{Introduction}

\indent
Epidemic dynamics on complex networks have recently attracted an increasing amount of attention from researchers
since Barab\'{a}si and Albert~\cite{1} proposed a scale-free network model, in which the
degree distribution $p(k)$ follows a power-law distribution ($p(k)\propto k^{-\gamma}$, where $\gamma$ usually
ranges between 2 and 3). The epidemic spreading systems (e.g., infectious diseases, computer viruses, rumor,
information diffusion etc.) can be modeled~\cite{2,3,4,5,6}---in such a way that the host population is
modeled as a contact network where nodes stand for individuals and each edge connecting two nodes describes potential
contact between two individuals.

According to the propagation characteristics of diseases and classical epidemic models,
many network epidemic models have been developed and used to obtain a lot of useful and insightful results~\cite{7,8,9,10}.
For those disease where infected individuals will not obtain lifelong immunity and
can return to susceptible state immediately (e.g., encephalitis, influenza, gonorrhea).
Pastor-Satorras and Vespignani presented the SIS model in highly heterogeneous networks
(i.e., scale-free networks)~\cite{7}. The most striking result is that they found
 the absence of the epidemic threshold in these networks. That is,
  the threshold approaches zero in the limit of a large number of edges and nodes, and even quite a
small infectious rate can produce a major epidemic outbreak.
When considering those diseases that can lead to permanent immunity and people are never infected
by that disease again (e.g., the parotitis, measles and SARS, etc.),
Moreno et al. showed a detailed analytical and numerical study on SIR epidemic model in scale-free networks~\cite{11}
and also found similar conclusions with the SIS network model.
These results have inspired a great number of related works~\cite{8,13,14,15,17,23},
and most of them suggest that both the properties of diseases and the network topology
determine the dynamical behavior of the spread of epidemics.

Epidemic diseases (cholera, tuberculosis, influenza, Ebola, etc.)
continue to have both a major impact on human beings and economic cost to society
now even after the development of modern medicine.
Therefore, any gain in understanding the dynamics and control of epidemic transmission
has potential for significant impact---and hence has attracted much attention from scientific comments.
In particular, vaccination and quarantine (isolation) are two important factors for preventing and controlling epidemic outbreak.
In order to study the role of these two controls, vaccination and quarantine were introduced into mathematical compartmental models.

On one hand, we know that the vaccination strategy will prevent (or reduce) the spread of many human diseases by vaccination of susceptible individuals.
However, there is clear evidence that some vaccines are not completely effective, namely, vaccines rarely cover the entire population
and only provide finite-time immunity against infection~\cite{17,18,19,20,21,23,26,27,28}.
Kribs-Zaleta and Velasco-Hern$\acute{a}$ndez~\cite{17} added vaccination into SIS model,
they studied rich dynamical behaviors, such as backward bifurcation and bistability.
Li et al.\cite{21} considered a two-dimensional SVIS model that vaccinated individuals
become susceptible again when vaccine loses its protective properties.
They exhibited backward bifurcation under certain conditions on treatment.
Peng et al.~\cite{23} found that the effective vaccination can linearly decrease the epidemic prevalence in small-world
networks. Moreover, it can act exponentially for scale-free networks.
Liu et al. \cite{26} showed that improvement of the efficiency of vaccines can weaken the necessary condition for disease eradication
by studying two SVIR models that describe continuous vaccination and pulse vaccination strategies.
Geng et al. \cite{27} investigated a discrete multi-group SVIR epidemic model with imperfect vaccination, and proved the global asymptotic stability of equilibria.
In~\cite{28} the vaccination into age structure of the
host population is considered to study Hep.B transmission.
The cost-effective balance of interventions methods by optimal control theory is determined.

On the other hand, since many diseases are transmitted from infectious to susceptible individuals through social contacts,
 an epidemic can be controlled  by isolating infected individuals.
 Therefore, quarantine is also a natural and widely practised method of human disease control.
Eastwood~\cite{29} showed that spontaneous quarantine in H{1}N{1} pandemic has a great impact on reducing the final size of the epidemic.
In~\cite{30} the SIR model in the presence of quarantine is analyzed, in which individuals alter their local neighborhoods with constant
quarantine probability. They found a phase transition at a critical rewiring (quarantine) threshold above which the epidemic is stopped from spreading.
In  \cite{31}, the authors discussed the application of optimal and sub-optimal controls to SARS. They demonstrated
 that the early quarantine and isolation strategies are critically important to control the outbreaks of epidemics. Otherwise, the control
effect will be much worse.
Li et al. \cite{32} proposed an SIQRS epidemic model on scale-free networks.
They found that the epidemic threshold significantly depends on the topology of complex networks and quarantine rate.
Then, in~\cite{33} an SIQRS epidemic model with demographics and quarantine
on complex heterogeneous networks is investigated, and the global epidemic behavior is analyzed.

By setting up a good epidemic model and thoroughly understanding it,
we can have many advantages of preventing invasion of infection to the population.
However, previous studies of mathematical models incorporating vaccination
ignore either the population structure, imperfect vaccination, quarantine
(isolation) of the symptomatic individuals, or demography (birth and death).
This paper aims to provide a systematic framework that couples public vaccination
and quarantine on scale-free network.
For this purpose, we perform two different cases as follows:

\noindent \textbf{Case (1)} In order to study these diseases spreading through population that lead to permanent natural immunity,
such as parotitis, measles and SARS, etc.
We develop a general network-based SVIQR model by extending the compartmental SVIR model by Liu et al.~\cite{26}.
\begin{equation}\label{eq11}\left\{\begin{array}{l}
 \frac{dS(t)}{dt}=b-\lambda S(t)I(t)-(\mu+d)S(t),\\
 \frac{dV(t)}{dt}=\mu S(t)-\delta\lambda V(t)I(t)-(d+\alpha)V(t),\\
 \frac{dI(t)}{dt}=\lambda S(t)I(t)+\delta\lambda V(t)I(t)-(\gamma+d)I(t),\\
 \frac{dR(t)}{dt}= \gamma I(t)+\alpha V(t)-d R(t).\\
 \end{array}\right.
\end{equation}

\noindent \textbf{Case (2)} For investigating those diseases that infected individuals will not obtain lifelong immunity
that people may be infected by that disease again (e.g., influenza, gonorrhea).
We introduce a network-based SVIQS model according to the compartmental SIV model by Kribs-Zaleta and Velasco-Hern\'{a}ndez~\cite{17}.
\begin{equation}\label{eq12}\left\{\begin{array}{l}
 \frac{dS(t)}{dt}=bN-\lambda S(t)\frac{I(t)}{N}+\gamma I(t)+\omega V(t)-(\mu+d)S(t),\\
 \frac{dV(t)}{dt}=\mu S(t)-\delta\lambda V(t)\frac{I(t)}{N}-(d+\omega)(t),\\
 \frac{dI(t)}{dt}=\lambda S(t)\frac{I(t)}{N}+\delta\lambda V(t)\frac{I(t)}{N}-(\gamma+d)I(t).\\
 \end{array}\right.
\end{equation}
$S(t), V(t), I(t)$, $R(t)$ denote the number of susceptible, vaccinated, infectious and recovered individuals at time $t$, respectively.
All coefficients are assumed to be positive and the biological interpretation is listed in Table 1.

In the mathematical theory of infectious diseases control, there are two important approaches to get theoretical
insights about how an infectious disease may be managed, reduced and possibly eradicated: the qualitative analysis of
mathematical models and the optimal control theory.
In this paper, however, we pay particular attention to mathematical compartmental theory of models in the presence of
imperfect vaccination and quarantine on scale-free networks.
The rest of this paper is organized as follows: In Sect.2, we first describe stochastic
evolution mechanism of  models and formulate two models based on mean-field theory.
Second, we introduce some preliminaries which is useful for the main results in Sect.3.
The main results are in Sect.4, we obtain the  basic reproduction number of two models.
By constructing corresponding Lyapunov functions, we analyze the
global stability of the disease-free and endemic equilibria of SVIQR model.
Then, we give qualitative analysis of stability in SVIQS model by using a monotone iterative technique.
In Sect.5, some numerical simulations are conducted to approve analytic results and to show the influence of
imperfect vaccination and quarantine in the control of disease spread.
Finally, a brief discussion is given in Sect.6 to conclude the paper.


\section{Description and formation of epidemic models}

\subsection{Stochastic model of SIR model with imperfect vaccination and quarantine}
In order to study \textbf{Case (1)}: the disease spreads through the population that lead to permanent natural immunity,
such as parotitis, measles and SARS, etc. We develop an SVIQR model on scale-free networks.

To simulate the process of interaction, a complex network $N$ is
established and individuals are spatially distributed on this network, where each node of $N$ is either vacant or occupied by one
individual. The nodes are enumerated with index $i = 1,2,\cdots,N$. The degree $k_{i}$ of node $i$ is the
number of links between node $i$ and other nodes. We divide all nodes into six categories: susceptible ($S$),
vaccinated ($V$), infected ($I$), quarantined ($Q$), recovered ($R$) and vacant ($O$).
\begin{table}[!h]
 \renewcommand\arraystretch{0.8}
  \caption{Notation used in models}\label{table-1}
  \vspace{0.5mm}
  \begin{tabular}{cl}
   \hline
   \rowcolor{cyan}
  Symbol & Meaning\\
  \hline
  $p(k)$, $n$     & Proportion of nodes with degree $k$. The max degree.\\
   \hdashline[1pt/1pt]
  $\langle k\rangle$  &The average degree of a vertices in the network~$\left(\langle k\rangle=\sum_{k}k p(k)\right)$.\\
   \hdashline[1pt/1pt]
  $b$, $d$        & The birth rate. The natural death rate. \\
  \hdashline[1pt/1pt]
  $\lambda$ & The transmission rate following one link. $\lambda(k)$ is the degree-dependent infection rate.\\
 \hdashline[1pt/1pt]
  $\mu_{k}$  &Vaccination rate of susceptible nodes with degree $k$.\\
 \hdashline[1pt/1pt]
  $\omega$  &Relapse rate of vaccinated individuals~(i.e., each vaccinated returns to being susceptible \\
            &after an average time period of $\frac{1}{\omega}$ due to temporary immunity). \\
 \hdashline[1pt/1pt]
  $\alpha$ & The rate of the vaccinated immunity becoming to permanent natural immunity.\\
 \hdashline[1pt/1pt]
 $\delta$ & Denotes the degree to which the vaccine-induced protection against infection is inefficient.\\
 \hdashline[1pt/1pt]
  $\beta_{k}$  & The quarantine rate of infected nodes with degree $k$.\\
  \hdashline[1pt/1pt]
   $\gamma$, $\eta$    &The recovery rate of infected, quarantined nodes.\\
  \hdashline[1pt/1pt]
  $\varphi(k_{i})$  &The infectivity of infected node $i$ with degree $k_{i}$.\\
  \hline
  \end{tabular}
\end{table}
Our SVIQR model is based on four factors as follows and a schematic of
the model is shown in Fig.~1(a).

\textbf{(1) Demographic impact (birth and death)}: This observation mainly depends on
what type of disease is being modeled. For instance, seasonal influenza can
be modeled well in a population without demography. Conversely, for the human
immunodeficiency virus (HIV) where infection span decades, then the demographic
impact should be taken into account~\cite{9}. In this paper, we propose our model with birth and death of individuals,
which would be more reasonable and precise to analyze a long-lasting epidemic spreading in an open population.

\emph{\textbf{Birth $O \rightarrow S$}}: Each vacant node $i$ randomly selects a neighbor at each time step.
If the neighbor is non-vacant node, the vacant node $i$ will give birth to a new susceptible
node with birth rate $b$. Due to the physiological limitation, it is assumed that each non-vaccant node
 generates the same birth contacts $A$ at each time step. \\
 \begin{figure}[htbp]
\centering
\subfigure[]{\includegraphics[height=4.5cm,width=6.5cm]{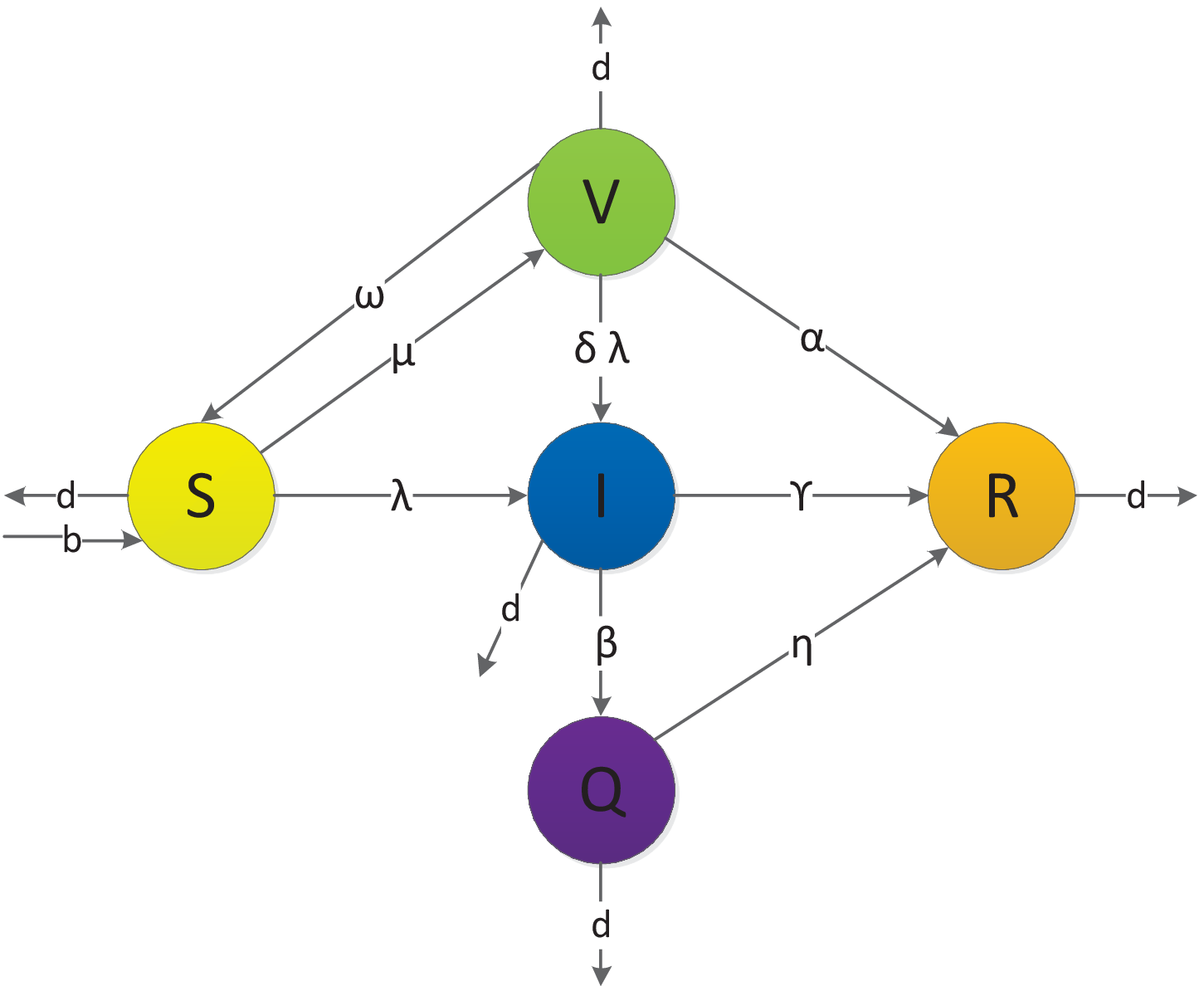}
\label{Fig1(a)}}
\subfigure[]{\includegraphics[height=4.5cm,width=6.5cm]{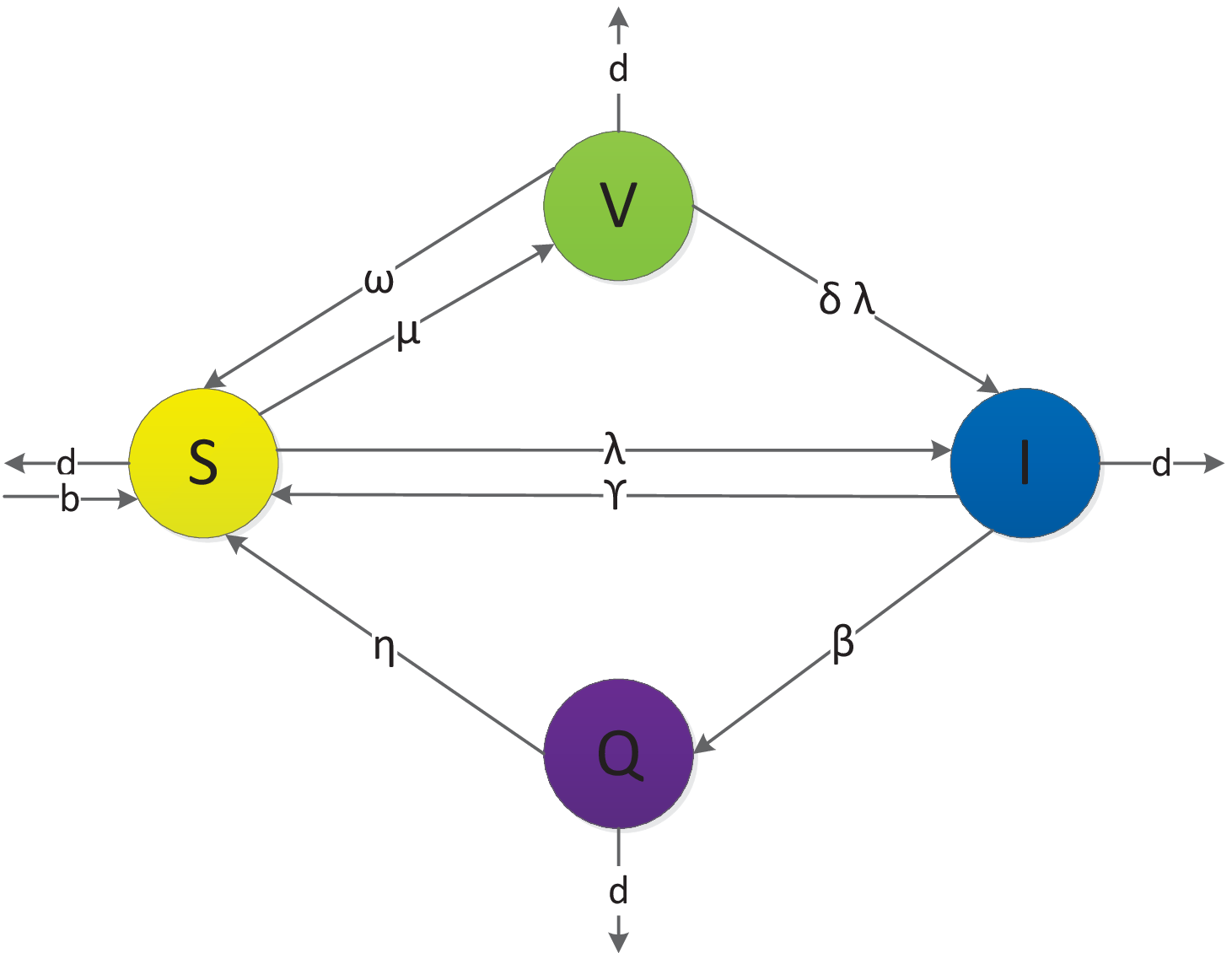}
\label{Fig1(b)}}
\caption{\footnotesize{(a). Flow chart of the SVIQR model;
~~~~~~~~(b). Flow chart of the SVIQS model.}}
\label{Fig1}
\end{figure}

 \emph{\textbf{Death $S/V/I/Q/R\rightarrow O$}}: All individuals occupied nodes die with death rate $d$ at
each time step. If a non-vacant node dies, there is an vacant node left.

\textbf{(2) SIR epidemic framework}: The model has been widely used in modeling disease spread. Each individual is assumed to have
one of three states: susceptible (S), infected (I), recovery (R).
It contains two process:

\emph{\textbf{Infection $S\rightarrow I$}}: To start the spreading
process, a few nodes are chosen as infected nodes. Each infected node $i$ contacts
each of its susceptible neighbors with the probability $\frac{\varphi(k_{i})}{k_{i}}$ at each time step.
If an infected node $i$ has contact with one of its susceptible neighbors $j$,
then this susceptible neighbor will be infected with probability $\lambda$.

\emph{\textbf{Recovery $I\rightarrow R$}}: All infected nodes can be cured and become recovery with rate $\gamma$ at each time step.

 \textbf{(3) Vaccination:} $S \rightarrow V\rightarrow R$: Vaccination operates by vaccinating susceptible individuals with degree $k$
with vaccination rate $\mu_{k}$ at each time step.
At the same time, some vaccinated-immunity individuals ($V$) may obtain natural immunity ($R$) with a conversion rate $\alpha$.

\emph{\textbf{Imperfect Vaccination $I \leftarrow V \rightarrow S$}}:
In reality, some vaccines may offer temporary immunity
and only provide finite-time immunity against infection. Therefore, the vaccinated individual
return to susceptible state with resusceptibility rate $\omega$ as vaccine wears off.
 Meanwhile, if one vaccinated node connects to an infected node following one edge,
this vaccinated individuals may be infected with probability $\delta\lambda$.

\textbf{(4) Quarantine: $I\rightarrow Q\rightarrow R$}.
At each time step, the infected individuals with degree $k$ will be quarantined with probability $\beta_{k}$. Meanwhile, the quarantined
individuals will recover to recovered state with rate $\eta$.

\subsection{Mean-field model of SIR model with imperfect vaccination and quarantine}

 Suppose that $S_{k}(t)$,$V_{k}(t)$, $I_{k}(t)$, $Q_{k}(t)$, $R_{k}(t)$ denote the densities of susceptible,
vaccinated, quarantined, infected and recovered individuals with degree $k$ at time $t$ on scale-free network, respectively.
Let $N_{k}(t)=S_{k}(t)+V_{k}(t)+I_{k}(t)+Q_{k}(t)+R_{k}(t)$ be the total density of non-vacant nodes with degree $k$.
Thus, the density of vacant nodes with degree $k$ is $1-N_{k}(t)$.
Based on above hypotheses and notations, the dynamical mean-field equations are written as:
\begin{equation}\label{eq33}\left\{\begin{array}{l}
 \frac{dS_{k}(t)}{dt}=bk[1-N_{k}(t)]\Phi(t)-\lambda(k)S_{k}(t)\Theta(t)+\omega V_{k}(t)-(\mu_{k}+d)S_{k}(t),\\
 \frac{dV_{k}(t)}{dt}=\mu_{k}S_{k}(t)-\delta\lambda(k)V_{k}(t)\Theta(t)-(d+\omega+\alpha)V_{k}(t),\\
 \frac{dI_{k}(t)}{dt}=\lambda(k)S_{k}(t)\Theta(t)+\delta\lambda(k)V_{k}(t)\Theta(t)-(\gamma+\beta_{k}+d)I_{k}(t),\\
 \frac{dQ_{k}(t)}{dt}= \beta_{k} I_{k}(t)-(\eta+d) Q_{k}(t),\\
\frac{dR_{k}(t)}{dt}=\gamma I_{k}(t)+\eta Q_{k}(t)+\alpha V_{k}(t)-dR_{k}(t).
 \end{array}\right.
\end{equation}
where $\Theta(t)=\sum_{i=1}^n p(i|k)\frac{\varphi(i)}{i}I_{i}(t)$, ~$\Phi(t)=\sum_{i=1}^n p(i|k)\frac{A}{i}N_{i}(t)$.
The meaning of them will be interpreted in the following section.

\subsection{Stochastic model of SIS model with imperfect vaccination and quarantine}

For investigating \textbf{Case (2)}: for those diseases where the infected individuals will return to susceptible state immediately(e.g., influenza, gonorrhea).
Similarly, we consider an SVIQS model on the same scale-free network,
and divide all nodes into five categories: susceptible(S), vaccinated (V), infected (I), quarantined(Q) and vacant (O).
The process of SVIQS model is showed in Fig.~1(b) and also includes following four factors:

\textbf{(1) Demographic impact (birth and death)}: This process is the same with SVIQR model in above section.

\textbf{(2) SIS epidemic framework}: Each individual is simply assumed to have only
two states: susceptible (S), infected (I). At each time step, the infection process is similar to SIR model.
But the recovery process is that all infected nodes can be cured and become susceptible with recovery rate $\gamma$.

\textbf{(3) Vaccination and Imperfect vaccination}: The only difference of this process with SVIQR model
  is that vaccinated immunity cannot become permanent natural immunity.

\textbf{(4) Quarantine: $I\rightarrow Q\rightarrow S$}. At each time step, the infected individuals with degree $k$ also be quarantined
 with probability $\beta_{k}$. Meanwhile, quarantined individuals recover to susceptible state with a recovery rate $\eta$.

\subsection{Mean-field model of SIS model with imperfect vaccination and quarantine}

Based on above assumptions, similarly,
$N_{k}(t)=S_{k}(t)+V_{k}(t)+I_{k}(t)+Q_{k}(t)$ is the total density of non-vacant nodes with degree $k$ at time $t$.
We obtain the following mean-field equations of SVIQS model:
\begin{equation}\label{eq21}\left\{\begin{array}{l}
 \frac{dS_{k}(t)}{dt}=bk[1-N_{k}(t)]\Phi(t)-\lambda(k)S_{k}(t)\Theta(t)+\gamma I_{k}(t)+\eta Q_{k}(t)+\omega V_{k}(t)-(\mu_{k}+d)S_{k}(t),\\
 \frac{dV_{k}(t)}{dt}=\mu_{k}S_{k}(t)-\delta\lambda(k)V_{k}(t)\Theta(t)-(d+\omega)V_{k}(t),\\
 \frac{dI_{k}(t)}{dt}=\lambda(k)S_{k}(t)\Theta(t)+\delta\lambda(k)V_{k}(t)\Theta(t)-(\gamma+\beta_{k}+d)I_{k}(t),\\
 \frac{dQ_{k}(t)}{dt}= \beta_{k}I_{k}(t)-(\eta+d) Q_{k}(t).\\
 \end{array}\right.
\end{equation}
where $\Theta(t)=\sum_{i=1}^n p(i|k)\frac{\varphi(i)}{i}I_{i}(t)$, ~$\Phi(t)=\sum_{i=1}^n p(i|k)\frac{A}{i}N_{i}(t)$.

The meanings of the parameters and variables in model ~\eqref{eq33} and ~\eqref{eq21} are as follows:

1. The expression $p(i|k)$ is the probability that a node of degree $k$ is connected to a node of degree $i$.
In the present paper, we primarily study epidemic transmission on uncorrelated networks,
the probability is considered independent of the connectivity of the node from which the link is emanating.
Therefore, $P(i|k)=\frac{iP(i)}{\langle k\rangle}$.

2. The function $\Theta(t)=\sum_{i=1}^n p(i|k)\frac{\varphi(i)}{i}I_{i}(t)$~describes the probability
 of a link pointing to an infected individual. We note that $\varphi(k)$ denotes the average number of edges from which a node with degree $k$ can transmit the disease.
It should be noted that various types of the infectivity $\varphi(k)$ have been
studied, such as $\varphi(k)=k$ \cite{11,8,15,23}, $\varphi(k)$=A \cite{37}, $\varphi(k)=k^{m}$ \cite{38},
$\varphi(k)=\frac{ak^{m}}{1+bk^{m}}$ \cite{36}.
On uncorrelated networks, $\Theta(t)=\frac{1}{\langle k\rangle}\sum_{i=1}^n \varphi(i)p(i)I_{i}(t)$.

3. The function $\Phi(t)=\sum_{i=1}^n p(i|k)\frac{A}{i}N_{i}(t)$ is the probability of
fertile contact between nodes with degree $k$ and its neighbors with degree $i$. The factor
$\frac{1}{i}$ accounts for the probability that one of the neighboring individual of a vacant
node with degree $i$ will activate this vacant node at the present time step.
It is assumed that, at each time step, every individual generates the same birth contacts $A$, here $A=1$.
Therefore, $bk[1-N_{k}(t)]\Phi(t)$ represents density of new born individuals per unit time.
On uncorrelated networks, $\Phi(t)=\frac{1}{\langle k\rangle}\sum_{i=1}^n p(i)N_{i}(t).$

4. Let $S(t)=\sum^{n}_{k=1}S_{k}(t)p(k)$ is the average density of susceptible individuals.
we similarly define $V(t)$,  $I(t)$, $Q(t)$, $R(t)$ are the average density of vaccinated, infected,
quarantined, recovered individuals, respectively.

\begin{rmk}
\textbf{(1)} We consider diseases that cannot cause vertical infection~(an infection caused by pathogens that uses mother-to-child transmission) and ignore disease-related deaths.
Meanwhile, we ignore the time for vaccinees to obtain immunity and infected people to be quarantined.
\textbf{(2)} In our models, we extend the constant vaccination rate and quarantine rate to be degree-related,
 which is realistic when we consider the reasonable control strategies related to degree $k$ on contact networks.
 \textbf{(3)} Systems \eqref{eq33} and \eqref{eq21} are two general models with vaccination and quarantine.
When there is no vaccination ($\mu_{i}=0$), systems are network-based SIQR and SIQS models.
When $\omega=0$, system \eqref{eq33} can depict the corresponding network models of \eqref{eq11} in \cite{26}.
 In addition, when there is no quarantine, the models become network-based SVIR and SVIS models with imperfect vaccination.
 \end{rmk}

\section{Preliminaries}

 In later section, we will consider the stability of systems \eqref{eq33} and \eqref{eq21}, which is one of
 the most important topics in the study of epidemiology. Now let us state some preliminaries which are needed later.
\begin{lem}[\cite{39}]
When $t\geq0$, $x(0)\geq0$, if $f>0, g>0$ and $\frac{dx(t)}{dt}\geq f-gx$, we have $\liminf\limits_{t\rightarrow{\infty}}x(t)\geq\frac{f}{g}$.
While, if $\frac{dx(t)}{dt}\leq f-gx$, we have $\limsup\limits_{t\rightarrow{\infty}}x(t)\leq\frac{f}{g}$.
\end{lem}
\begin{lem}[\cite{40}]
(i).~If $ W(t)$ is a Lyapunov function of system $X'=TX$, $X(0)=x_{0}$;~~
(ii).~x(n) is a solution of this system bounded for all $n\geq0$,
 then there is a number $c$ such that $x(n)\rightarrow M\bigcap W^{-1}(c)$ as $n\rightarrow\infty$.
 where $M$ is the largest invariant set in $G=\{x| W'(t)=0\}$ and $ W^{-1}(c)=\{x|W(x)=c, x\in R^{m}\}$.
 \end{lem}

From system~\eqref{eq33} and \eqref{eq21}, we show the evolutions of $N_{k}(t)$ are governed by the differential equation:
\begin{equation}\label{eq22}
\frac{dN_{k}(t)}{dt}=bk[1-N_{k}(t)]\Phi(t)-dN_{k}(t).
\end{equation}
Zhu et al. \cite{34} drew the following conclusions:\\
(1) When $b~\leq~d$, $\lim\limits_{t \rightarrow \infty}N_{k}(t)=0$. The population becomes extinct, there is no other dynamic behaviors;\\
(2) When $b>d$, $\lim\limits_{t \rightarrow \infty}N_{k}(t)=N^{*}_{k}$, where $N^{*}_{k}=\frac{bk\Phi^{*}}{d+bk\Phi^{*}}$, $\Phi^{*}=\frac{1}{\langle k\rangle}\sum_{i} \frac{ip(i)b\Phi^{*}}{d+bi\Phi^{*}}, k=1,2,\cdots,n.$

Since the original system and the limiting system have the same asymptotic dynamical behaviors, to study the stability
of systems~\eqref{eq33} and~\eqref{eq21}, we only need consider their limiting systems under
$N^{*}_{k}=S_{k}(t)+V_{k}(t)+I_{k}(t)+Q_{k}(t)+R_{k}(t)$ in \eqref{eq33} and $N^{*}_{k}=S_{k}(t)+V_{k}(t)+I_{k}(t)+Q_{k}(t)$ in \eqref{eq21}.
Based on above results, we only consider the case of $b>d$.
The limiting systems corresponding to systems~\eqref{eq33} and~\eqref{eq21} are written as follows:
\begin{equation}\label{eq34}\left\{\begin{array}{l}
 \frac{dS_{k}(t)}{dt}=bk[1-N^{*}_{k}]\Phi^{*}-\lambda(k)S_{k}(t)\Theta(t)+\omega V_{k}(t)-(\mu_{k}+d)S_{k}(t),\\
 \frac{dV_{k}(t)}{dt}=\mu_{k}S_{k}(t)-\delta\lambda(k)V_{k}(t)\Theta(t)-(d+\omega+\alpha)V_{k}(t),\\
 \frac{dI_{k}(t)}{dt}=\lambda(k)S_{k}(t)\Theta(t)+\delta\lambda(k)V_{k}(t)\Theta(t)-(\gamma+\beta_{k}+d)I_{k}(t),\\
 \frac{dQ_{k}(t)}{dt}= \beta_{k}I_{k}(t)-(\eta+d) Q_{k}(t),\\
\frac{dR_{k}(t)}{dt}=\gamma I_{k}(t)+\eta Q_{k}(t)+\alpha V_{k}(t)-dR_{k}(t).
 \end{array}\right.
\end{equation}
with initial conditions $0<S_{k}(0),V_{k}(0),I_{k}(0),Q_{k}(0),R_{k}(0)\leq N^{*}_{k}<1, k=1,2,\cdots,n.$
\begin{equation}\label{eq24}\left\{\begin{array}{l}
 \frac{dS_{k}(t)}{dt}=bk[1-N^{*}_{k}]\Phi^{*}-\lambda(k)S_{k}(t)\Theta(t)+\gamma I_{k}(t)+\eta Q_{k}(t)+\omega V_{k}(t)-(\mu_{k}+d)S_{k}(t),\\
 \frac{dV_{k}(t)}{dt}=\mu_{k}S_{k}(t)-\delta\lambda(k)V_{k}(t)\Theta(t)-(d+\omega)V_{k}(t),\\
 \frac{dI_{k}(t)}{dt}=\lambda(k)S_{k}(t)\Theta(t)+\delta\lambda(k)V_{k}(t)\Theta(t)-(\gamma+\beta_{k}+d)I_{k}(t),\\
 \frac{dQ_{k}(t)}{dt}= \beta_{k}I_{k}(t)-(\eta+d) Q_{k}(t).
 \end{array}\right.
\end{equation}
with initial conditions $0<S_{k}(0),V_{k}(0),I_{k}(0),Q_{k}(0)\leq N^{*}_{k}<1, k=1,2,\cdots,n.$

In order to ensure that the models are well-posed,
it is necessary to stress the positivity and boundedness of solutions of system~\eqref{eq33} and~\eqref{eq21}.
We note the following sets according to system~\eqref{eq33} and~\eqref{eq21},
$$\Omega_{R}=\{(S_{k}(t), V_{k}(t), I_{k}(t), Q_{k}(t), R_{k}(t))|0~\leq S_{k}(t), V_{k}(t), I_{k}(t),Q_{k}(t), R_{k}(t) \leq~N^{*}_{k}<1\};$$
$$\Omega_{S}=\{(S_{k}(t), V_{k}(t), I_{k}(t), Q_{k}(t))|0~\leq S_{k}(t), V_{k}(t), I_{k}(t),Q_{k}(t)\leq~N^{*}_{k}<1\}.$$
Then, we obtain the following Propositions to show the positivity and boundedness of solutions of SVIQR model and SVIQS model, respectively.
See~\cite{43} for the same method to prove these Propositions.
\begin{pro}
The set $\Omega_{R}$ must be positively invariant, that is,
if $(S_{k}(t), V_{k}(t), I_{k}(t), Q_{k}(t), R_{k}(t))_{k}$ is a solution of \eqref{eq34} satisfying initial conditions,
then $0\leq S_{k}(t), V_{k}(t), I_{k}(t),Q_{k}(t), R_{k}(t)\leq N^{*}_{k}$ for any $t$ and $k$.
\end{pro}
\begin{pro}
The set $\Omega_{S}$ must be positively invariant, that is,
if $(S_{k}(t), V_{k}(t), I_{k}(t), Q_{k}(t))_{k}$ is a solution of~\eqref{eq24} satisfying initial conditions,
then $0~\leq S_{k}(t), V_{k}(t), I_{k}(t), Q_{k}(t)\leq N^{*}_{k}$ for any $t$ and $k = 1,2,\cdots,n$.
\end{pro}

Since $\Omega_{R}$ and $\Omega_{S}$ are positively invariant absorbing sets.
It is sufficient to consider the dynamics
of the flow generated by system~\eqref{eq34} and~\eqref{eq24} in $\Omega_{R}$ and $\Omega_{S}$.

\section{Main results}
In this section, qualitative analysis of system~\eqref{eq34} and~\eqref{eq24} are presented,
including the basic reproduction number, the global stability of
disease-free and endemic equilibria. We develop different mathematical methods
to study the global dynamics of these two systems.
For simplicity, we let $\beta_{k}=\beta$ as a constant quarantine rate,
and note $\xi=\gamma+\beta+d$, $f=\frac{\beta}{\eta+d}$ and $\Lambda_{k}=bk[1-N^{*}_{k}]\Phi^{*}$ in the following analysis.
\subsection{The equilibria and basic reproduction number }
The basic reproduction number $R_{0}$ is the average number of new infections generated
 by a single newly infectious individual during the full infectious period.
In epidemiology, for most cases, it is the most common threshold parameter,
associated with bifurcation phenomena.
In the case of the simplest epidemic models,
$R_{0}$ is proved to be a sharp threshold parameter, completely
determining the global dynamics of system \cite{41}. That is to say,
if $R_{0}<1$, the disease-free equilibrium is global asymptotically stable; If $R_{0}>1$,
however, a unique endemic equilibrium is global asymptotically stable. Namely,
prototypical $R_{0}$ threshold behavior features a `forward' bifurcation, in which the unique endemic
equilibrium exists only for $R_{0}>1$.
In systems exhibiting a backward bifurcation, however, the endemic equilibrium exists for $R_{0}<1$,
so that under certain initial conditions it is possible for an invasion to succeed with $R_{0}<1$ \cite{17}.
Moreover, reducing $R_{0}$ back below one would not eradicate disease.
In this section, we draw this important threshold parameters and study equilibria of models.

\subsubsection{Analysis of SVIQR epidemic  model}
 It is easy to see that all biologically feasible equilibria of system~\eqref{eq34} are admitted by the following equation,
\begin{equation}\label{eq35}\left\{\begin{array}{l}
\Lambda_{k}-\lambda(k)S_{k}\Theta+\omega V_{k}-(\mu_{k}+d)S_{k}=0,\\
\mu_{k}S_{k}-\delta\lambda(k)V_{k}\Theta-(d+\omega+\alpha)V_{k}=0,\\
\lambda(k)S_{k}\Theta+\delta\lambda(k)V_{k}\Theta-(\gamma+\beta+d)I_{k}=0,\\
\beta I_{k}-(\eta+d) Q_{k}=0.
\end{array}\right.
\end{equation}
One can easily see that there always exists disease-free equilibrium $E^{0}=(S^{0}_{k}, V^{0}_{k},0, 0, R^{0}_{k})$ of ~\eqref{eq34}, where
\begin{small}
$$ S^{0}_{k}=\frac{\Lambda_{k}(d+\omega+\alpha)}{(d+\omega+\alpha)d+(d+\alpha)\mu_{k}},~~
V^{0}_{k}=\frac{\Lambda_{k}\mu_{k}}{(d+\omega+\alpha)d+(d+\alpha)\mu_{k}},~~I^{0}_{k}= Q^{0}_{k}=0, R^{0}_{k}=\frac{\alpha}{d}V^{0}_{k}.$$
\end{small}
and any positive equilibrium $E^{*}=(S^{*}_{k}, V^{*}_{k},I^{*}_{k},Q^{*}_{k}, R^{*}_{k})(k=1,2,\cdots,n$) satisfies that
\begin{small}
$$ S^{*}_{k}=\frac{\Lambda_{k}(d+\omega+\alpha+\delta\lambda(k)\Theta^{*})}{(d+\omega+\alpha+\delta\lambda(k)\Theta^{*})(d+\lambda(k)\Theta^{*})+(d+\alpha+\delta\lambda(k)\Theta^{*})\mu_{k}},$$
$$V^{*}_{k}=\frac{\Lambda_{k}\mu_{k}}{(d+\omega+\alpha+\delta\lambda(k)\Theta^{*})(d+\lambda(k)\Theta^{*})+(d+\alpha+\delta\lambda(k)\Theta^{*})\mu_{k}},~~
R^{*}_{k}=\frac{\alpha V^{*}_{k}+\gamma I^{*}_{k}+\eta Q^{*}_{k}}{d},$$
$$ I^{*}_{k}=\frac{\lambda(k)\Theta^{*}}{(\gamma+\beta+d)}\frac{\Lambda_{k}(d+\omega+\alpha+\delta\lambda(k)\Theta^{*}+\delta \mu_{k})}{(d+\omega+\alpha+\delta\lambda(k)\Theta^{*})(d+\lambda(k)\Theta^{*})+(d+\alpha+\delta\lambda(k)\Theta^{*})\mu_{k}},
~Q^{*}_{k}=\frac{\beta}{\eta+d}I^{*}_{k},~~ \Theta^{*}=\frac{1}{k}\sum_{i=1}^n \varphi(i)p(i)I^{*}_{i}.$$
\end{small}
\begin{thm}
Define the basic reproduction number
$$R_{0}= \frac{1}{\langle k\rangle}\sum_{k=1}^n \varphi(k) p(k)\lambda(k)\frac{bk(1-N^{*}_{k})\Phi^{*}(d+\omega+\alpha+\delta \mu_{k})}{(\gamma+\beta+d)[(d+\alpha+\omega)d+(d+\alpha)\mu_{k}]}.$$
There always exists a disease-free equilibrium $E^{0}=(S^{0}_{k}, V^{0}_{k}, R^{0}_{k}, 0,0)_{k}$.
~If and only if $R_{0}>1$, system~\eqref{eq34} has a unique endemic equilibrium point
$E^{*}=(S^{*}_{k}, V^{*}_{k},I^{*}_{k},Q^{*}_{k},R^{*}_{k})$,  $k=1,2,\cdots,n$.
\end{thm}
\textbf{Proof.}  We follow ~\eqref{eq35} and obtain
\begin{small}\begin{equation}\label{eq36}
I_{k}=\frac{\lambda(k)\Theta}{\xi}\frac{\Lambda_{k}(d+\omega+\alpha+\delta\lambda(k)\Theta+\delta\mu_{k})}
{(d+\omega+\alpha+\delta\lambda(k)\Theta)(d+\lambda(k)\Theta)+(d+\alpha+\delta\lambda(k)\Theta)\mu_{k}}.
\end{equation}
\end{small}
Substituting~\eqref{eq36} into $\Theta(t)=\frac{1}{\langle k\rangle}\sum_{i=1}^n \varphi(i) p(i)I_{i}(t)$.
Then, we obtain the self-consistency equation:
\begin{small}$$\Theta=\frac{1}{\langle k\rangle}\sum_{k=1}^n \varphi(k) p(k)\frac{\lambda(k)\Theta}{\xi}\frac{\Lambda_{k}(d+\omega+\alpha+\delta\lambda(k)\Theta+\delta\mu_{k})}
{(d+\omega+\alpha+\delta\lambda(k)\Theta)(d+\lambda(k)\Theta)+(d+\alpha+\delta\lambda(k)\Theta)\mu_{k}}.$$
\end{small}
We note $F(\Theta)=\Theta\left(1-f(\Theta)\right)=0$, where
$$f(\Theta)= \frac{1}{\langle k\rangle}\sum_{k=1}^n \varphi(k) p(k)\frac{\lambda(k)}{\xi}\frac{\Lambda_{k}(d+\omega+\alpha+\delta\lambda(k)\Theta+\delta \mu_{k})}{A_{k}\Theta^{2}+B_{k}\Theta+C_{k}}.$$
and~$A_{k}=\delta\lambda^{2}(k)$,~~$B_{k}=(\delta d+\omega+d+\alpha+\delta\mu_{k})\lambda(k)$,~$C_{k}=(d+\alpha+\omega)d+(d+\alpha)\mu_{k}.$ Then, we obtain
$$f'(\Theta)= \frac{1}{\langle k\rangle}\sum_{k=1}^n \frac{\varphi(k) p(k)\Lambda_{k}\lambda(k)}{\xi}\left(\frac{-\delta\lambda(k)A_{k}\Theta^{2}-2A_{k}\Theta(d+\omega+\alpha+\delta \mu_{k})-\Delta_{k}}{\left(A_{k}\Theta^{2}+B_{k}\Theta+C_{k}\right)^{2}}\right).$$
where $\Delta_{k}=\lambda(k)[\delta^{2}\mu_{k}d+\delta\mu_{k}(\omega+\delta\mu_{k})+(\omega+d+\alpha)(\omega+d+\alpha+\delta\mu_{k})].$
We get $f'(\Theta)<0$ for $\Theta\geq0$, $\lim\limits_{\Theta\rightarrow +\infty}f(\Theta)=0$. Thus, $F(\Theta)=0$ has a nontrivial solution on interval$(0,1)$ if and only if $f(0)>1$,
which yields
$$\frac{1}{\langle k\rangle}\sum_{k=1}^n \varphi(k) p(k)\lambda(k)\frac{bk(1-N^{*}_{k})\Phi^{*}(d+\omega+\alpha+\delta \mu_{k})}{(\gamma+\beta+d)[(d+\alpha+\omega)d+(d+\alpha)\mu_{k}]}>1.$$
Therefore, the unique endemic equilibrium exists only for $E^{*}$ if $R_{0}>1$ in system~\eqref{eq34}. 
That completes the proof.
\subsubsection{Analysis of SVIQS epidemic model}
For system \eqref{eq24}, with $N^{*}_{k}=S_{k}(t)+V_{k}(t)+I_{k}(t)+Q_{k}(t)$, we get the equilibria should satisfy
\begin{equation}\label{eq26}\left\{\begin{array}{l}
 \mu_{k}(N^{*}_{k}-V_{k}(t)-I_{k}(t)-Q_{k}(t))-\delta\lambda(k)V_{k}(t)\Theta(t)-(d+\omega)V_{k}(t)=0,\\
 \lambda(k)(N^{*}_{k}-V_{k}(t)-I_{k}(t)-Q_{k}(t))\Theta(t)+\delta\lambda(k)V_{k}(t)\Theta(t)-(\gamma+\beta+d)I_{k}(t)=0,\\
 \beta I_{k}(t)-(\eta+d) Q_{k}(t)=0.
 \end{array}\right.
\end{equation}
It is easy to see that system \eqref{eq24} has a disease-free equilibrium $E^{0}: S_{k}=\frac{(d+\omega)N^{*}_{k}}{\mu_{k}+\omega+d}, V_{k}=\frac{\mu_{k}N^{*}_{k}}{\mu_{k}+\omega+d}, I_{k}=Q_{k}=0$.\\
Additionally, the endemic equilibrium $E^{*}=(S^{*}_{k}, V^{*}_{k},I^{*}_{k},Q^{*}_{k}),k=1,2,\cdots,n,$ which satisfies that
\begin{small}
$$S^{*}_{k}=\frac{(\Lambda_{k}+(\gamma+\eta f)I^{*}_{k})(\omega+d+\delta\lambda(k)\Theta^{*})}{(\omega+d+\delta\lambda(k)\Theta^{*})(\lambda(k)\Theta^{*}+d)+\mu_{k}(d+\delta\lambda(k)\Theta^{*})},
~~~V^{*}_{k}=\frac{(\Lambda_{k}+(\gamma+\eta f)I^{*}_{k})\mu_{k}}{(\omega+d+\delta\lambda(k)\Theta^{*})(\lambda(k)\Theta^{*}+d)+\mu_{k}(d+\delta\lambda(k)\Theta^{*})},$$
$$I^{*}_{k}=\frac{\lambda(k)N^{*}_{k}\Theta^{*}(\omega+\delta\lambda(k)\Theta^{*}+\delta\mu_{k}+d)}{\xi
(\omega+\delta\lambda(k)\Theta^{*}+\mu_{k}+d)+\lambda(k)\Theta^{*}(1+f)(\omega+\delta\lambda(k)\Theta^{*}+\delta\mu_{k}+d)},
~~~~Q^{*}_{k}=\frac{\beta}{\eta+d}I^{*}_{k},~~~ \Theta^{*}=\frac{1}{k}\sum_{i=1}^n \varphi(i)p(i)I^{*}_{i}.$$
\end{small}

Let us first consider the disease-free equilibrium $E^{0}$. We can calculate the basic reproductive
number for the disease-which determines the local stability of $E^{0}$ by means of the
Jacobian matrix. When vaccination and quarantine are present, we obtain the basic reproductive number
\begin{small}
$$R_{0}=\frac{1}{\langle k\rangle}\sum_{i=1}^n\frac{\lambda(i)\varphi(i)p(i)bi(1-N^{*}_{i})\Phi^{*}(\omega+d+\delta\mu_{i})}{d(\gamma+\beta+d)(\omega+\mu_{i}+d)}.$$
\end{small}
Note that $\widetilde{R}_{0}=R_{0}\mid_{\mu_{k}=0}=\frac{1}{\langle k\rangle}\sum_{i=1}^n\frac{\lambda(i)\varphi(i)p(i)bi(1-N^{*}_{i})\Phi^{*}}{d(\gamma+\beta+d)}$
. Obviously, $R_{0}\leq \widetilde{R}_{0}$ for all $\mu_{k}\geq0$ is satisfied, which depicts the influence of vaccination on $R_{0}$.

Next, we investigate the endemic equilibria $E^{*}$ and then obtain the following self-consistency equation of $\Theta$,
\begin{small}
$$\Theta=\frac{1}{\langle k\rangle}\sum_{k=1}^n \varphi(k)p(k)\frac{\lambda(k)N^{*}_{k}\Theta(\omega+\delta\lambda(k)\Theta+\delta\mu_{k}+d )}
{\xi(\omega+\delta\lambda(k)\Theta+\mu_{k}+d )+\lambda(k)\Theta(1+f)(\omega+\delta\lambda(k)\Theta+\delta\mu_{k}+d)}\triangleq\Theta f(\Theta).$$
\end{small}
It is equivalent to $\Theta\left(1-f(\Theta)\right)=0$, where
$$f(\Theta)=\frac{1}{\langle k\rangle}\sum_{k=1}^n \varphi(k)p(k)\frac{\lambda(k)N^{*}_{k}(\omega+\delta\lambda(k)\Theta+\delta\mu_{k}+d)}
{A_{k}\Theta^{2}+B_{k}\Theta+C_{k}}.$$
and $A_{k}=\delta\lambda^{2}(k)(1+f)$,~$B_{k}=\delta\lambda(k)\xi+\lambda(k)(1+f)(d+\omega+\delta\mu_{k})$
,~~$C_{k}=(d+\mu_{k}+\omega)\xi$.

Obviously, the positive solutions can be determined by $f(\Theta)=1$ and $f(0)=R_{0}$. Furthermore, we have
\begin{small}
$$f'(\Theta)=\frac{1}{\langle k\rangle}\sum_{k=1}^n \lambda(k)\varphi(k)p(k)N^{*}_{k} \left( \frac{-\delta\lambda(k)A_{k}\Theta^{2}-2(\omega+\delta\mu_{k}+d)A_{k}\Theta+\delta\lambda(k)C_{k}-(\omega+\delta\mu_{k}+d)B_{k}}{[A_{k}\Theta^{2}+B_{k}\Theta+C_{k}]^{2}}\right),$$
\end{small}
and $\lim\limits_{\Theta \rightarrow +\infty}f(\Theta)=0$, $f(\Theta)>0$. We note
\begin{small}$h(\Theta)\triangleq-\delta\lambda(k)A_{k}\Theta^{2}-2(\omega+\delta\mu_{k}+d)A_{k}\Theta+\delta\lambda(k)C_{k}-(\omega+\delta\mu_{k}+d)B_{k}.$\end{small}
We consider the special cases to study the qualitative behavior of $f(\Theta)=1$ based on the sign of $f'(\Theta)$ as follows:

\noindent \textbf{Case 1:} when $\delta\lambda(k)C_{k}-(\omega+\delta\mu_{k}+d)B_{k}\leq0$, which is equivalent to
$$R^{1}_{0}(\mu_{k})\triangleq \frac{\delta(1-\delta)\xi\mu_{k}}{(1+f)(\omega+d+\delta\mu_{k})^{2}}\leq1.$$
Here, $f'(\Theta)<0$ for $\Theta\geq0$. Obviously, the equation $f(\Theta)=1$ has a unique positive solution only if $f(0)=R_{0}>1$.
 Therefore, if $R^{1}_{0}(\mu_{k})\leq1$, the system exhibit 'forward' bifurcation.

\noindent \textbf{Case 2:} when $\delta\lambda(k)C_{k}-(\omega+\delta\mu_{k}+d)B_{k}>0$ i.e. $R^{1}_{0}(\mu_{k})>1$.
Hence, $h(\Theta)=0$ has two solutions,
\begin{small}
 $$\Theta_{1}=\frac{-(\omega+d+\delta\mu_{k})-\sqrt{\delta(1-\delta)\xi\mu_{k}/(1+f)}}{\delta\lambda(k)}<0, ~~ \Theta_{2}=\frac{-(\omega+d+\delta\mu_{k})+\sqrt{\delta(1-\delta)\xi\mu_{k}/(1+f)}}{\delta\lambda(k)}>0.$$
 \end{small}
Then, we get $h(\Theta)>0$ and $f'(\Theta)>0$ for $\Theta\in[0,\Theta_{2}]$ and  $h(\Theta)<0$, $f'(\Theta)<0$ for $\Theta\in[\Theta_{2}, +\infty]$.\\
 Let
\begin{small}
$$R^{2}_{0}(\mu_{k})\triangleq f(\Theta_{2})=\frac{1}{\langle k\rangle}\sum_{k=1}^n \frac{\delta\varphi(k)p(k)\lambda(k)N^{*}_{k}}
{2\sqrt{\delta(1-\delta)\xi\mu_{k}(1+f)}-(\omega+d+\delta\mu_{k})(1+f)+\xi \delta}.$$
\end{small}
Obviously, $R^{2}_{0}(\mu_{k})>0$. When $R^{1}_{0}(\mu_{k})>1$, we obtain the following results:\\
If $R_{0}>1$ or $R^{2}_{0}(\mu_{k})=1$ or $R_{0}=1$, equation $f(\Theta)=1$ only has one positive solution and system~\eqref{eq24} has a unique endemic equilibrium.\\
If $R_{0}<1<R^{2}_{0}(\mu_{k})$, the equation $f(\Theta)=1$ has two positive solutions, then \eqref{eq24} has multiple endemic equilibria.\\
If $R^{2}_{0}(\mu_{k})<1$, there is no positive solution.

From above analysis, we therefore rewrite these conclusions as follows:
\begin{thm}
Consider \eqref{eq24},
the following statements hold:\\
(1) There always exists a disease-free equilibrium $E^{0}=\left(\frac{(d+\omega)N^{*}_{k}}{\mu_{k}+\omega+d}, \frac{\mu_{k}N^{*}_{k}}{\mu_{k}+\omega+d}, 0, 0\right)_{k}$, $k=1,2,\cdots,n$. \\
(2) A unique endemic equilibrium exists if $R_{0}>1.$ \\
(3) A unique endemic equilibrium exists if $R^{1}_{0}(\mu_{k})>1$=$R_{0}$ or $R^{1}_{0}(\mu_{k})>1$=$R^{2}_{0}(\mu_{k})$ is satisfied.\\
(4) There exists at least two endemic equilibria if $R_{0}<1<\min\{R^{1}_{0}(\mu_{k}),R^{2}_{0}(\mu_{k})\}$.\\
(5) There is no endemic equilibria if  $R^{1}_{0}(\mu_{k})\leq1$ and $R_{0}<1$ or $R^{2}_{0}(\mu_{k})<1<R^{1}_{0}(\mu_{k})$.
\end{thm}
\begin{rmk}
For two special cases $\delta=0$ and $\delta=1$ respectively representing completely effective and utterly useless vaccination,
we find that systems \eqref{eq34} and \eqref{eq24} all show 'forward' bifurcation.
For general condition $0<\delta\ll1$, Theorem~4.1 shows that \eqref{eq34} only undergoes a forward bifurcation.
We remark that Theorem~4.2 indicates the possibility of multiple endemic equilibria for $R_{0}<1$ and hence the potential
occurrence of a backward bifurcation in \eqref{eq24}. However, if $R^{1}_{0}(\mu_{k})\leq1$ holds, there is no endemic equilibrium for $R_{0}<1$
 that \eqref{eq24} only undergoes forward bifurcation.
 \end{rmk}


\subsection{Stability analysis of equilibria of SVIQR model}
It is important to analyze the stability of equilibria, as it will indicate whether the disease will
die out eventually, or it will become endemic. In particular, the global stability of epidemiological
model becomes much more interesting from realistic views to theoretical views.
 In this section, we will establish the global stability of disease-free and
endemic equilibria of system~\eqref{eq34}. They are achieved by using Lyapunov functional approach.
\subsubsection{Stability analysis of the disease-free equilibrium (DFE)}
 Firstly, we prove the DFE $E^{0}$ is locally asymptotically
stable. We claim this by analyzing the Jacobian matrix of model~\eqref{eq34}
evaluating at $E^{0}$. The same method of proof is showed in \cite{26}, we omit the proof here.
\begin{thm} If $R_{0}<1$, then the DFE of system \eqref{eq34} is locally asymptotically
stable and unstable if~$R_{0}>1$.
\end{thm}

Next, we consider the global attractivity of $E^{0}$ by construct suitable Lyapunov function,
 which is one of the most popular used methods in the study of global stability.
\begin{thm}~~If $R_{0}<1$, then the disease-free
equilibrium $E_0$  is globally asymptotically
stable. If $R_{0}>1$, then the disease-free equilibrium is unstable.
\end{thm}
\textbf{Proof.} From Proposition 3.1,
we prove that $E^{0}$ is globally asymptotically stable in positive invariant $\Omega_{R}$.
Using the fact that $g(x)=x-1-lnx\geq g(1)=0$ for all $x>0$,
 we consider the following Lyapunov function
$$W(t)=\frac{1}{\langle k\rangle}\sum^{n}_{k=1}\varphi(k)p(k)S^{0}_{k}\left(\frac{S_{k}}{S^{0}_{k}}-1-ln\frac{S_{k}}{S^{0}_{k}}\right)+\frac{1}{\langle k\rangle}\sum^{n}_{k=1}\varphi(k)p(k)V^{0}_{k}\left(\frac{V_{k}}{V^{0}_{k}}-1-ln\frac{V_{k}}{V^{0}_{k}}\right)+\Theta(t).$$
The Lyapunov function $W(t)$ is non-negative and is defined with respect to disease-free equilibrium
$E_{0}$, which is a global minimum. According to \eqref{eq35}, the equilibrium satisfies
$$\Lambda_{k}=-\omega V^{0}_{k}+(\mu_{k}+d)S^{0}_{k},~~~\mu_{k}S^{0}_{k}=(d+\omega+\alpha)V^{0}_{k}.$$

Differentiating $W(t)$ along solutions to \eqref{eq34} gives:
\begin{small}\begin{eqnarray*}
\frac{d W(t)}{dt}&=&\frac{1}{\langle k\rangle}\sum^{n}_{k=1}\varphi(k)p(k)\left(1-\frac{S^{0}_{k}}{S_{k}}\right)\frac{d S_{k}}{dt}+\frac{1}{\langle k\rangle}\sum^{n}_{k=1}\varphi(k)p(k)\left(1-\frac{V^{0}_{k}}{V_{k}}\right)\frac{d V_{k}}{dt}+\frac{1}{\langle k\rangle}\sum^{n}_{k=1}\varphi(k)p(k)\frac{d I_{k}}{dt}\\
&=&\frac{1}{\langle k\rangle}\sum^{n}_{k=1}\varphi(k)p(k)\omega V^{0}_{k}\left(2-\frac{S^{0}_{k}V_{k}}{S_{k}V^{0}_{k}}-\frac{S_{k}V^{0}_{k}}{S^{0}_{k}V_{k}}\right)+\frac{1}{\langle k\rangle}\sum^{n}_{k=1}\varphi(k)p(k)dS^{0}_{k}\left(2-\frac{S_{k}}{S^{0}_{k}}-\frac{S^{0}_{k}}{S_{k}}\right)\\
& & +\frac{1}{\langle k\rangle}\sum^{n}_{k=1}\varphi(k)p(k)(d+\alpha)V^{0}_{k}\left(3-\frac{V_{k}}{V^{0}_{k}}-\frac{S^{0}_{k}}{S_{k}}-\frac{S_{k}V^{0}_{k}}{S^{0}_{k}V_{k}}\right)\\
& & +\left[\frac{1}{\langle k\rangle}\sum^{n}_{k=1}\varphi(k)p(k)\lambda(k)\frac{\left(S^{0}_{k}+\delta V^{0}_{k}\right)}{\gamma+\beta+d}-1\right](\gamma+\beta+d)\Theta(t).
\end{eqnarray*}
\end{small}
Due to the fact that arithmetic mean is greater than or equal to the geometric mean, we have the following inequalities
$$2-\frac{S_{k}}{S^{0}_{k}}-\frac{S^{0}_{k}}{S_{k}}\leq~0,~~~2-\frac{S^{0}_{k}V_{k}}{S_{k}V^{0}_{k}}-\frac{S_{k}V^{0}_{k}}{S^{0}_{k}V_{k}}\leq0,$$
\begin{small}\begin{eqnarray*}
3-\frac{V_{k}}{V^{0}_{k}}-\frac{S^{0}_{k}}{S_{k}}-\frac{S_{k}V^{0}_{k}}{S^{0}_{k}V_{k}}=-g\left(\frac{V_{k}}{V^{0}_{k}}\right)-g\left(\frac{S^{0}_{k}}{S_{k}}\right)
-g\left(\frac{S_{k}V^{0}_{k}}{S^{0}_{k}V_{k}}\right)\leq~0,
\end{eqnarray*}
\end{small}
$$\left[\frac{1}{\langle k\rangle}\sum^{n}_{k=1}\varphi(k)p(k)\lambda(k)\frac{\left(S^{0}_{k}+\delta V^{0}_{k}\right)}{\gamma+\beta+d}-1\right]=R_{0}-1.$$
Thus, $R_{0}\leq 1$ ensures that $\frac{d W(t)}{dt}\leq0$ holds.
Each solution of system~\eqref{eq34} tends to $\Gamma_{1}$, where $\Gamma_{1}$ is the largest invariant
subset satisfied $\frac{d W(t)}{dt}=0$. Note that when $R_{0}<1$, the equality holds only if $S_{k}(t)=S^{0}_{k}$, $V_{k}(t)=V^{0}_{k}$.
Hence we obtain $\Gamma_{1}$=$\{E_{0}\}$, the global stability of $E_{0}$ follows by Lemma 3.2 .
This completes the proof.

\subsubsection{Global stability of endemic equilibrium }
Now, we are in the position to state
the global stability of endemic equilibrium $E^{*}$ of system~~\eqref{eq34}.
\begin{thm}
If $R_{0}>1$, then the endemic equilibrium $E^{*}$ of~\eqref{eq34} is globally asymptotically stable.
\end{thm}
\textbf{Proof.} We construct a Lyapunov function $W(t)=W_{S}(t)+W_{V}(t)+W_{I}(t)$, where
\begin{small}$$W_{S}(t)=\frac{1}{\langle k\rangle}\sum^{n}_{k=1}\varphi(k)p(k)S^{*}_{k}\left(\frac{S_{k}}{S^{*}_{k}}-1-ln\frac{S_{k}}{S^{*}_{k}}\right),$$
$$W_{V}(t)=\frac{1}{\langle k\rangle}\sum^{n}_{k=1}\varphi(k)p(k)V^{*}_{k}\left(\frac{V_{k}}{V^{*}_{k}}-1-ln\frac{V_{k}}{V^{*}_{k}}\right),~~~
W_{I}(t)=\Theta^{*}\left(\frac{\Theta}{\Theta^{*}}-1-ln\frac{\Theta}{\Theta^{*}}\right).$$
\end{small}
According to system~\eqref{eq35}, the positive equilibrium satisfies,
\begin{small}$$\Lambda_{k}=\lambda(k)S^{*}_{k}\Theta^{*}-\omega V^{*}_{k}+(\mu_{k}+d)S^{*}_{k}, ~~~\mu_{k}S^{*}_{k}=\delta\lambda(k)V^{*}_{k}\Theta^{*}+(d+\omega+\alpha)V^{*}_{k},$$
$$(\gamma+\beta+d)=\frac{1}{\langle k\rangle}\sum^{n}_{k=1}\varphi(k)p(k)\lambda(k)[S^{*}_{k}+\delta V^{*}_{k}].$$
\end{small}
Differentiating $W_{S}(t), W_{V}(t), W_{I}(t)$ along the solution of system~\eqref{eq35}, we obtain
\begin{small}\begin{eqnarray*}
\frac{d W_{S}(t)}{dt}&=&\frac{1}{\langle k\rangle}\sum^{n}_{k=1}\varphi(k)p(k)\left(1-\frac{S^{*}_{k}}{S_{k}}\right)\frac{d S_{k}}{dt}\\
&=& \frac{1}{\langle k\rangle}\sum^{n}_{k=1}\varphi(k)p(k)\left(1-\frac{S^{*}_{k}}{S_{k}}\right)\left(\Lambda_{k}-\lambda(k)S_{k}\Theta(t)+\omega V_{k}-(\mu_{k}+d)S_{k}\right)\\
&=& \frac{1}{\langle k\rangle}\sum^{n}_{k=1}\varphi(k)p(k)\left(1-\frac{S^{*}_{k}}{S_{k}}\right)\left[\lambda(k)(S^{*}_{k}\Theta^{*}-S_{k}\Theta)+\omega (V_{k}-V^{*}_{k})+
(\mu_{k}+d)(S^{*}_{k}-S_{k}(t))\right]\\
&=&\frac{1}{\langle k\rangle}\sum^{n}_{k=1}\varphi(k)p(k)\lambda(k)\left(1-\frac{S^{*}_{k}}{S_{k}}\right)(S^{*}_{k}\Theta^{*}-S_{k}\Theta(t))+
\frac{1}{\langle k\rangle}\sum^{n}_{k=1}\varphi(k)p(k)\omega V^{*}_{k}\\
& &\left(\frac{V_{k}}{V^{*}_{k}}-\frac{S^{*}_{k}V_{k}}{S_{k}V^{*}_{k}}-1+ \frac{S^{*}_{k}}{S_{k}}\right)+\frac{1}{\langle k\rangle}\sum^{n}_{k=1}\varphi(k)p(k)(\mu_{k}+d)S^{*}_{k}\left(2-\frac{S_{k}}{S^{*}_{k}}-\frac{S^{*}_{k}}{S_{k}}\right).
\end{eqnarray*}\end{small}
\begin{small}\begin{eqnarray*}
\frac{d W_{V}(t)}{dt}&=&\frac{1}{\langle k\rangle}\sum^{n}_{k=1}\varphi(k)p(k)\left(1-\frac{V^{*}_{k}}{V_{k}}\right)\frac{d V_{k}}{dt}\\
&=& \frac{1}{\langle k\rangle}\sum^{n}_{k=1}\varphi(k)p(k)\left(1-\frac{V^{*}_{k}}{V_{k}}\right)\left(\mu_{k}S_{k}-\delta\lambda(k)V_{k}\Theta-(d+\omega+\alpha)V_{k}\right)\\
&=& \frac{1}{\langle k\rangle}\sum^{n}_{k=1}\varphi(k)p(k)\left(1-\frac{V^{*}_{k}}{V_{k}}\right)\left[\mu_{k}S^{*}_{k}(\frac{S_{k}}{S^{*}_{k}}-1)+\delta\lambda(k)(V^{*}_{k}\Theta^{*}-V_{k}\Theta)
+(d+\omega+\alpha)(V^{*}_{k}-V_{k})\right]\\
&=&\frac{1}{\langle k\rangle}\sum^{n}_{k=1}\varphi(k)p(k)\delta\lambda(k)\left(1-\frac{V^{*}_{k}}{V_{k}}\right)(V^{*}_{k}\Theta^{*}-V_{k}\Theta)+\frac{1}{\langle k\rangle}\sum^{n}_{k=1}\varphi(k)p(k)\mu_{k}S^{*}_{k}\left(\frac{V^{*}_{k}}{V_{k}}-\frac{S_{k}V^{*}_{k}}{S^{*}_{k}V_{k}}-1+ \frac{S_{k}}{S^{*}_{k}}\right)\\
& &+\frac{1}{\langle k\rangle}\sum^{n}_{k=1}\varphi(k)p(k)(\omega+d+\alpha)V^{*}_{k}\left(2-\frac{V_{k}}{V^{*}_{k}}-\frac{V^{*}_{k}}{V_{k}}\right).
\end{eqnarray*}\end{small}
\begin{small}\begin{eqnarray*}
\frac{d W_{I}(t)}{dt}&=&\left(1-\frac{\Theta^{*}}{\Theta}\right)\frac{1}{\langle k\rangle}\sum^{n}_{k=1}\varphi(k)p(k)\frac{d I_{k}}{dt}\\
&=& \frac{1}{\langle k\rangle}\sum^{n}_{k=1}\varphi(k)p(k)\lambda(k)\left(S_{k}+\delta V_{k}\right)\left(\Theta-\Theta^{*}\right)-(\gamma+\beta+d)\left(\Theta-\Theta^{*}\right)\\
&=& \frac{1}{\langle k\rangle}\sum^{n}_{k=1}\varphi(k)p(k)\lambda(k)\left(S_{k}-S^{*}_{k}\right) \left(\Theta-\Theta^{*}\right)+\frac{1}{\langle k\rangle}\sum^{n}_{k=1}\varphi(k)p(k)\delta\lambda(k)\left(V_{k}-V^{*}_{k}\right) \left(\Theta-\Theta^{*}\right).
\end{eqnarray*}\end{small}
Combining the derivatives of $W_{S}(t), W_{V}(t), W_{I}(t)$ yields
\begin{small}\begin{eqnarray*}
\frac{d W(t)}{dt}&=&\frac{1}{\langle k\rangle}\sum^{n}_{k=1}\varphi(k)p(k)\lambda(k)S^{*}_{k}\Theta^{*}\left(2-\frac{S_{k}}{S^{*}_{k}}-\frac{S^{*}_{k}}{S_{k}}\right)
                    +\frac{1}{\langle k\rangle}\sum^{n}_{k=1}\varphi(k)p(k)\delta\lambda(k)V^{*}_{k}\Theta^{*}\\
                 & &\left(3-\frac{V_{k}}{V^{*}_{k}}-\frac{S^{*}_{k}}{S_{k}}-\frac{S_{k}V^{*}_{k}}{S^{*}_{k}V_{k}}\right)
                    +\frac{1}{\langle k\rangle}\sum^{n}_{k=1}\varphi(k)p(k)dS^{*}_{k}\left(2-\frac{S_{k}}{S^{*}_{k}}-\frac{S^{*}_{k}}{S_{k}}\right)\\
                 & &+\frac{1}{\langle k\rangle}\sum^{n}_{k=1}\varphi(k)p(k)(d+\alpha)V^{*}_{k}\left(3-\frac{V_{k}}{V^{*}_{k}}-\frac{S^{*}_{k}}{S_{k}}-\frac{S_{k}V^{*}_{k}}{S^{*}_{k}V_{k}}\right)\\
                 & &+\frac{1}{\langle k\rangle}\sum^{n}_{k=1}\varphi(k)p(k)\omega V^{*}_{k}\left(2-\frac{S_{k}V^{*}_{k}}{S^{*}_{k}V_{k}}-\frac{S^{*}_{k}V_{k}}{S_{k}V^{*}_{k}}\right).
\end{eqnarray*}
\end{small}
Since
\begin{small}
$$2-\frac{S_{k}}{S^{*}_{k}}-\frac{S^{*}_{k}}{S_{k}}\leq~0,~~~2-\frac{S_{k}V^{*}_{k}}{S^{*}_{k}V_{k}}-\frac{S^{*}_{k}V_{k}}{S_{k}V^{*}_{k}}\leq0,~~
3-\frac{V_{k}}{V^{*}_{k}}-\frac{S^{*}_{k}}{S_{k}}-\frac{S_{k}V^{*}_{k}}{S^{*}_{k}V_{k}}\leq~0.$$
\end{small}
and the equality holds if and only if $S_{k}=S^{*}_{k}$ and $V_{k}=V^{*}_{k}, k=1,2,\cdots,n.$

Let $\Gamma_{2}$ be the largest invariant set of $\{(S_{k}, V_{k}, I_{k}, Q_{k}, R_{k})|\frac{d W(t)}{dt}=0\}$.
Obviously, $\{E^{*}\}\subset \Gamma_{2}$. Now, $\frac{d W(t)}{dt}=0$ if and only if
 $S_{k}=S^{*}_{k}$ , $V_{k}=V^{*}_{k}$,  $I_{k}=I^{*}_{k}$,  $Q_{k}=Q^{*}_{k}$, $R_{k}=R^{*}_{k}$.
This proves that  $\{E^{*}\}\supset\Gamma$. Therefore,  $\{E^{*}\}=\Gamma_{2}$ and
it follows that $ \{E^{*}\}$ is globally asymptotically stable.
This completes the proof.

\subsection{Stability analysis of equilibria of SVIQS model}
We know that when forward bifurcation occurs, the condition $R_{0}<1$
is usually both necessary and sufficient for disease eradication, whereas it is no
longer sufficient when a backward bifurcation occurs.
As mentioned in Theorem 4.2, when $R^{1}_{0}(\mu_{k})\leq 1$, no endemic equilibrium exists for $R_{0}<1$, and there exists a unique
endemic equilibrium for $R_{0}>1$. That is, the condition $R^{1}_{0}(\mu_{k})\leq 1$ ensure the occurrence of a forward bifurcation in system~\eqref{eq24}.
In this section, we investigate the global stability of endemic equilibrium of system~\eqref{eq24} under the condition $R^{1}_{0}(\mu_{k})\leq 1$.
The main verifying tools are the comparison theorem and iteration principle.
\subsubsection{Stability analysis of the  disease-free equilibrium (DFE)}
We claim the local stability of DFE of system~\eqref{eq24} and omit the proof (the same as the proof of Theorem 4.3).
\begin{thm}~If $R_{0}<1$, the DFE of system~\eqref{eq24} is locally asymptotically
stable, and unstable while $R_{0}>1$.
\end{thm}
Next, we analyze the global stability of $E^{0}$ and obtain the following Theorem.
\begin{thm}~~If $\widetilde{R}_{0}<1$,~ then the disease-free
equilibrium $E^{0}$ is globally asymptotically
stable.
\end{thm}
\textbf{Proof.} Proposition 3.2 shows that $0 \leq S_{k}(t), V_{k}(t), I_{k}(t), Q_{k}(t)\leq N^{*}_{k}$ for any initial condition of system~\eqref{eq24}.
We firstly claim that $\lim\limits_{t\rightarrow +\infty}I_{k}(t)=0$ and $\lim\limits_{t\rightarrow +\infty}Q_{k}(t)=0$ for any $k=1,2,...,n$. We have
\begin{small}
$$ \frac{d \Theta(t)}{dt}\leq\frac{1}{\langle k\rangle}\sum^{n}_{k=1}\varphi(k)p(k)\lambda(k)\left(N^{*}_{k}-(1-\delta)V_{k}(t)\right)\Theta(t)-(\gamma+\beta+d)\Theta(t),$$
\end{small}
and $\Theta(t)\leq 
\Theta(0)e^{\left((\gamma+\beta+d)(\widetilde{R}_{0}-1)\right)t}$.
Since $\widetilde{R}_{0}<1$, we obtain $\lim\limits_{t\rightarrow +\infty}\Theta(t)=0$ and $\lim\limits_{t\rightarrow +\infty}I_{k}(t)=0$ for any $k$.
 Then, combining with the third equation of system ~\eqref{eq24}, we have
$\lim\limits_{t\rightarrow +\infty}Q_{k}(t)=0$. Therefore, there exists a small enough $\varepsilon>0$ to make sure that $0 \leq Q_{k}(t), I_{k}(t) \leq \varepsilon$.

Next, we show that $\lim\limits_{t\rightarrow +\infty}V_{k}(t)=V^{0}_{k}.$
According to system~\eqref{eq24} and Lemma 3.1, we derive that
 \begin{small}\begin{eqnarray*}
 \frac{d V_{k}(t)}{dt}
 &\leq & \mu_{k}N^{*}_{k}-(d+\omega+\mu_{k})V_{k}(t).
\end{eqnarray*}\end{small}
Therefore, $\limsup\limits_{t\rightarrow +\infty}V_{k}(t)\leq \frac{\mu_{k}N^{*}_{k}}{(d+\omega+\mu_{k})}=V^{0}_{k}$.

On the other hand, since $0 \leq Q_{k}(t), I_{k}(t) \leq \varepsilon$, we have
$$\frac{d V_{k}(t)}{dt}\geq\mu_{k}N^{*}_{k}-2\mu_{k}\varepsilon-(d+\omega+\mu_{k})V_{k}(t).$$
Setting $\varepsilon\rightarrow0$ and by Lemma 3.1, it follows that
 $\liminf\limits_{t\rightarrow +\infty} V_{k}(t)\geq\frac{\mu_{k}N^{*}_{k}}{d+\omega+\mu_{k}}=V^{0}_{k}.$

It is clear that $\lim\limits_{t\rightarrow +\infty} V_{k}(t)=\frac{\mu_{k}N^{*}_{k}}{d+\omega+\mu_{k}}=V^{0}_{k}.$
Since $S_{k}(t)=N^{*}_{k}-V_{k}(t)-I_{k}(t)-Q_{k}(t)$, it follows that $\lim\limits_{t\rightarrow +\infty}S_{k}(t)=S^{0}_{k}.$
 This proves that $E^{0}$ of system \eqref{eq24} is globally attractively for $\widetilde{R}_{0}<1$. Thus, combining with Theorem~4.6,
 $E_{0}$ of system ~\eqref{eq24} is globally  asymptotically stable. The proof is completed.
\begin{rmk}
The result of Theorem 4.7 can not exclude the stability of endemic equilibria which may exist for
$R_{0}<1$, as they should occur for parameter values which do not satisfy the theorem hypothesis.
Also, Theorem 4.6 and Theorem 4.7 also show that the vaccination extends the local (not necessarily global)
stability of the disease-free equilibrium.
\end{rmk}

In order to fill the gap that $\widetilde{R}_{0}<1$ doesn't preclude the stability of endemic equilibria for $R_{0}<1$,
we perform a global analysis under some assumption of parameters to make sure that $R^{1}_{0}(\mu_{k})\leq1$,
which has showed in Theorem 4.2 that there does not exist endemic equilibrium for $R_{0}<1$.
We observe that if $\omega\geq\eta=\gamma$ then $R^{1}_{0}(\mu_{k})\leq1$ is satisfied and the only steady state for $R_{0}<1$ is disease-free equilibrium.
We show the following global stability result, the method of proof is similar to Theorem~4.7, we omit the proof here.
\begin{thm}
Assume $\omega\geq\eta=\gamma$, if $R_{0}<1$, then the DFE of system \eqref{eq24} is globally asymptotically stable.
\end{thm}

\subsubsection{ Global stability of endemic equilibrium}

 Firstly, we analyze the uniform persistence of disease which can be used in the proof of global attractivity of endemic equilibrium of system~\eqref{eq24} .
 We discuss the permanence of disease and prove it using Thieme's result (Theorem 4.6) of \cite{42}. We define
 $$Y_{0}=\{(S_{k}(t),V_{k}(t),I_{k}(t),Q_{k}(t))\in \Omega_{S} | I(t)=\sum^{n}_{k=1}p(k)I_{k}>0\};~~\partial Y_{0}=\Omega_{S}\backslash Y_{0}.$$

We study system~\eqref{eq24} is uniform persistence on $(Y_{0},\partial Y_{0})$.
It is clear that $\Omega_{S}$ is positive invariant by proposition~3.2.
For any initial condition satisfied $\left(S_{k}(0),V_{k}(0),I_{k}(0),Q_{k}(0)\right)\in \partial Y_{0}$, we have
 $$\Theta'(t)=\frac{1}{\langle k\rangle}\sum^{n}_{k=1}\varphi(k)p(k)I'_{k}(t)\leq\frac{1}{\langle k\rangle}\sum^{n}_{k=1}\varphi(k)p(k) \lambda(k)N^{*}_{k}\Theta(t)(1+\delta)-(\gamma+\beta+d)\Theta(t),$$
and $\Theta(t)\leq \Theta(0)\exp{\left(\sum^{n}_{k=1}\varphi(k)p(k)\lambda(k)N^{*}_{k}(1+\delta)-(\gamma+\beta+d)\right)t}$.
Since $I(0)=0$, it is clear that $\Theta(0)=0$. Thus, $\Theta(t)=0$ and $\partial Y_{0}$ is the positive invariant.

If initial conditions satisfy $\left(S_{k}(0),V_{k}(0),I_{k}(0),Q_{k}(0)\right)\in Y_{0}$, then $I(0)>0$ for all $t>0$. Since
 $$I'(t)=\sum^{n}_{i=1}p(k)I'_{k}(t)\geq -\sum^{n}_{k=1}p(k)(\gamma+\beta+d)I_{k}(t)=-(\gamma+\beta+d)I(t).$$
We derive $I(t)>0$. Thus, $Y_{0}$ is also the positive invariant.
 Furthermore, there exists a compact set B in which all solutions of~\eqref{eq24} initiated in $Y_{0}$ will enter and remain forever after. The compactness condition
in \cite{26} is easily verified for set $B$.

We note that $\phi(t, X_{0})$ is the solution of system~\eqref{eq24} with the initial condition $X_{0}\in \Omega_{S}$,
 and $\omega(X_{0})$ is the $\omega$-limit of solution of system ~\eqref{eq24} with initial condition $X_{0}\in \Omega_{S}$.
 $$\Omega=\bigcup\left\{\omega(X_{0})|X_{0}\in \partial Y_{0}~and~\phi(t, X_{0}) \in \partial Y_{0}, for ~all~t>0\right\}.$$
 We consider system~\eqref{eq24} on $M_{\partial}=\left\{X_{0}|~\phi(t, X_{0}) \in \partial Y_{0},~\forall t\geq 0 \right\}$, and then obtain
 \begin{equation}\label{eq42}\left\{\begin{array}{l}
 \frac{dS_{k}(t)}{dt}=bk[1-N^{*}_{k}]\Phi^{*}+\eta Q_{k}(t)+\omega V_{k}(t)-(\mu_{k}+d)S_{k}(t),\\
 \frac{dV_{k}(t)}{dt}=\mu_{k}S_{k}(t)-(d+\omega)V_{k}(t),\\
 \frac{dI_{k}(t)}{dt}=-(\gamma+\beta+d)I_{k}(t),\\
 \frac{dQ_{k}(t)}{dt}= -(\eta+d) Q_{k}(t).
 \end{array}\right.
\end{equation}
Obviously, the disease-free equilibrium $E_{0}$ of~system~\eqref{eq24} is the only equilibrium of system~\eqref{eq42} in $M_{\partial}$.
We can easily obtain that $E_{0}$ is globally asymptotically stable in $M_{\partial}$.
Thus, $\Omega=\{E_{0}\}$. And $E_{0}$ is a covering of $\Omega$, which is isolated and is acyclic
(Because $E_{0}$ is the only equilibrium of system~\eqref{eq42} in $M_{\partial}$).

Finally, we state that the disease is uniformly persistent on $(Y_{0},\partial Y_{0})$ for $R_{0}>1$ based on above discussion.
The proof will be done if we show $E_{0}$ is weak repeller for $Y_{0}$, that is if $R_{0}>1$, then
$$\limsup\limits_{t \to +\infty}~d\left(\phi(t,X_{0}),~E_{0}\right)>0 ~for ~any~ X_{0}~\in~ Y_{0}.$$
Therefore, combining with $\Omega=\{E_{0}\}$, and $E_{0}$ is global stability in $M_{\partial}$,
we obtain the following theorem about the uniform persistence for $R_{0}>1$.
\begin{thm}~~If $R_{0}>1$,~then the disease is permanent on $(Y_{0},\partial Y_{0})$,i.e.,
there exists a constant $\varepsilon>0$, which is independent on the initial condition $X_{0}\in\Omega_{S}$, such that
$\liminf\limits_{t \to +\infty}~(I(t))>\varepsilon$ for any initial solution of system~\eqref{eq24}.
\end{thm}
\textbf{Proof.} We need only prove that $W^{s}(E_{0})\bigcap Y_{0}=\emptyset$, where
 $$W^{s}(E_{0})=\{X_{0}\in Y_{0}|\lim\limits_{t\to\infty}{\phi(t,X_{0})=E_{0}}\}.$$
If it is not true, we assume that there exists a solution $y\in Y_{0}$ of system \eqref{eq24} satisfied $\lim\limits_{t\to\infty}\Phi(y,X_{0})=E_{0}.$

Since
$R_{0}=\frac{1}{\langle k\rangle}\sum_{i=1}^n\frac{\lambda(i)\varphi(i)p(i)N^{*}_{i}(\omega+\delta\mu_{i}+d)}{(\gamma+\beta+d)(\omega+\mu_{i}+d)}>1$.
Then, there exists $\varepsilon>0$ to make sure that
$$R_{0}=\frac{1}{\langle k\rangle}\sum_{i=1}^n\frac{\lambda(i)\varphi(i)p(i)N^{*}_{i}}{\gamma+\beta+d}\left(\frac{\omega+\delta\mu_{i}+d}{\omega+\mu_{i}+d}-(1+\delta)\varepsilon\right)>1.$$

On the other hand, since $\phi(t, y) \to E_{0}$, for $\varepsilon>0$, there exists $T>0$ such that for $t>T$, we obtain
$$S^{0}_{k}-\varepsilon \leq S_{k}\leq S^{0}_{k}+\varepsilon ,
~ V^{0}_{k}-\varepsilon  \leq V_{k}\leq V^{0}_{k}+\varepsilon ,
~0 \leq I_{k}\leq \varepsilon ,~0 \leq Q_{k}\leq \varepsilon. $$
Next, we note $W(t)=\frac{1}{\langle k\rangle}\sum_{i=1}^n\frac{\varphi(i)p(i)}{\gamma+\beta+d}I_{i}(t)$, it follows that for all $t>T$,
\begin{eqnarray*}
 \frac{d W(t)}{dt}&=&\frac{1}{\langle k\rangle}\sum^{n}_{i=1}\frac{\varphi(i)p(i)}{\gamma+\beta+d}\left[\lambda(i)S_{i}(t)\Theta(t)+\delta\lambda(i)V_{i}(t)\Theta(t)-(\gamma+\beta+d)I_{i}(t)\right]\\
 &\geq&\frac{1}{\langle k\rangle}\sum^{n}_{i=1}\frac{\varphi(i)p(i)}{\gamma+\beta+d}\left\{\lambda(i)[S^{0}_{i}-\varepsilon+\delta(V^{0}_{i}-\varepsilon)]\Theta(t)-(\gamma+\beta+d)I_{i}(t)\right\}\\
 &=& \left[\frac{1}{\langle k\rangle}\sum_{i=1}^n\frac{\lambda(i)\varphi(i)p(i)N^{*}_{i}}{\gamma+\beta+d}\left(\frac{\omega+\delta\mu_{i}+d}{\omega+\mu_{i}+d}-(1+\delta)\varepsilon\right)-1\right](\gamma+\beta+d)W(t).
\end{eqnarray*}
Obviously, $\frac{d W(t)}{dt} >0$ and $W(0)> 0$, therefore  $W(t)\to +\infty$ as $t \to +\infty$. This contradicts to the boundedness of $W(t)$.
Hence, $W^{s}(E_{0})\bigcap Y_{0}=\emptyset$. This completes the proof.

Finally, assuming $\omega\geq\eta=\gamma$,
 we study the global stability of this endemic equilibrium $E^{*}$ of system~~\eqref{eq24}
  by constructing monotone iterative sequences which is similar to the method used in~\cite{43}.
 According to Theorem~4.9, it is clear that if $R_{0}>1$, the infection will always exist, that is,  $I(t)>\varepsilon$.
 More specific results are presented in the following theorem.
\begin{thm}
Suppose that $(S_{k}(t), V_{k}(t), I_{k}(t), Q_{k}(t))$ is the solution of~\eqref{eq24} satisfying initial condition. Assume that $\omega\geq\gamma=\eta$.
 If $R_{0}>1$,  then $\lim\limits_{t\rightarrow +\infty}S_{k}(t)=S^{*}_{k}$,$\lim\limits_{t\rightarrow +\infty}V_{k}(t)=V^{*}_{k}$, $\lim\limits_{t\rightarrow +\infty}I_{k}(t)=I^{*}_{k}$,
 $\lim\limits_{t\rightarrow +\infty}Q_{k}(t)=Q^{*}_{k}$, that is to say, the endemic equilibrium $E^{*}$ is globally attractive.
\end{thm}
\begin{rmk}
If $\omega\geq\gamma=\eta$, then we can also prove the global stability of equilibria of system~\eqref{eq24}
by constructing the Lyapunov function similar to the ones used in the proofs of Theorem~4.4 and Theorem~4.5.
 \end{rmk}
 \begin{rmk}
In our SVIQS model, we stress the special cases $\gamma=\eta=0$ or $\gamma=\eta=d$ in \cite{22}
 are two special cases that satisfy our general condition $R^{1}_{0}(\mu_{k})\leq1$ under which the system  only exhibits forward bifurcation.
 \end{rmk}

\section{Simulations}

We perform some simulations to illustrate and complement our theoretical results, and to
analyze the effects of parameters on the spread of epidemics, then find better
control strategies. Here, we mainly use model~\eqref{eq33} and~\eqref{eq21} to simulate the evolution
of epidemics. All the following simulations are based on a generated BA network
with network size $N=10000$,
this network evolves from initial network with size $m_{0}=4$
and adds each new node with $m = 3$ new edges.
The average degree of the generated network is $\langle k\rangle$=5.9988 and its maximum degree is $n = 236$.
\begin{figure}[h]
\centering
\tiny \subfigure[]{\includegraphics[height=4.5cm,width=7.2cm]{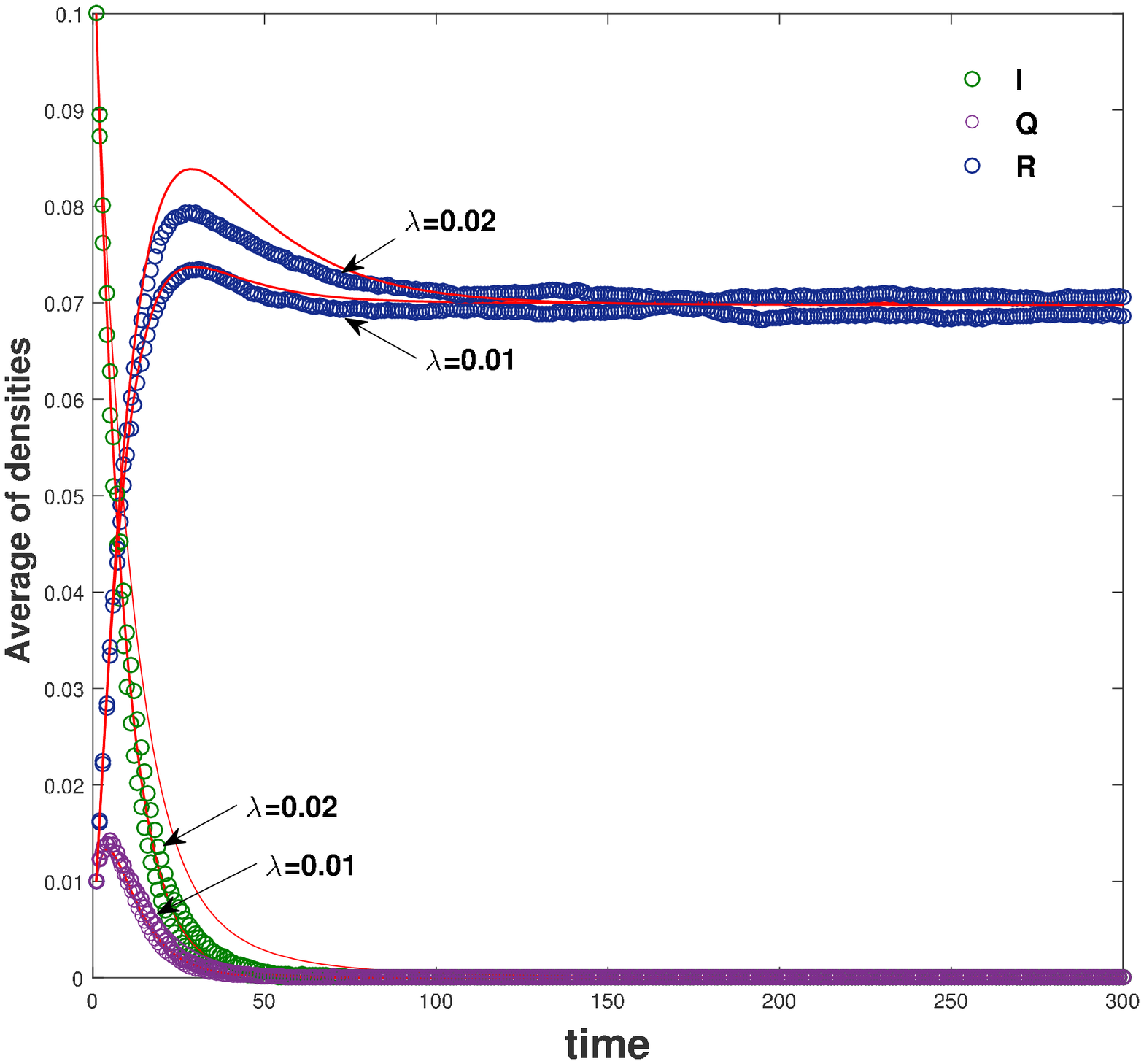}
\label{Fig2(a)}}
~~~~\tiny\subfigure[]{\includegraphics[height=4.5cm,width=7.2cm]{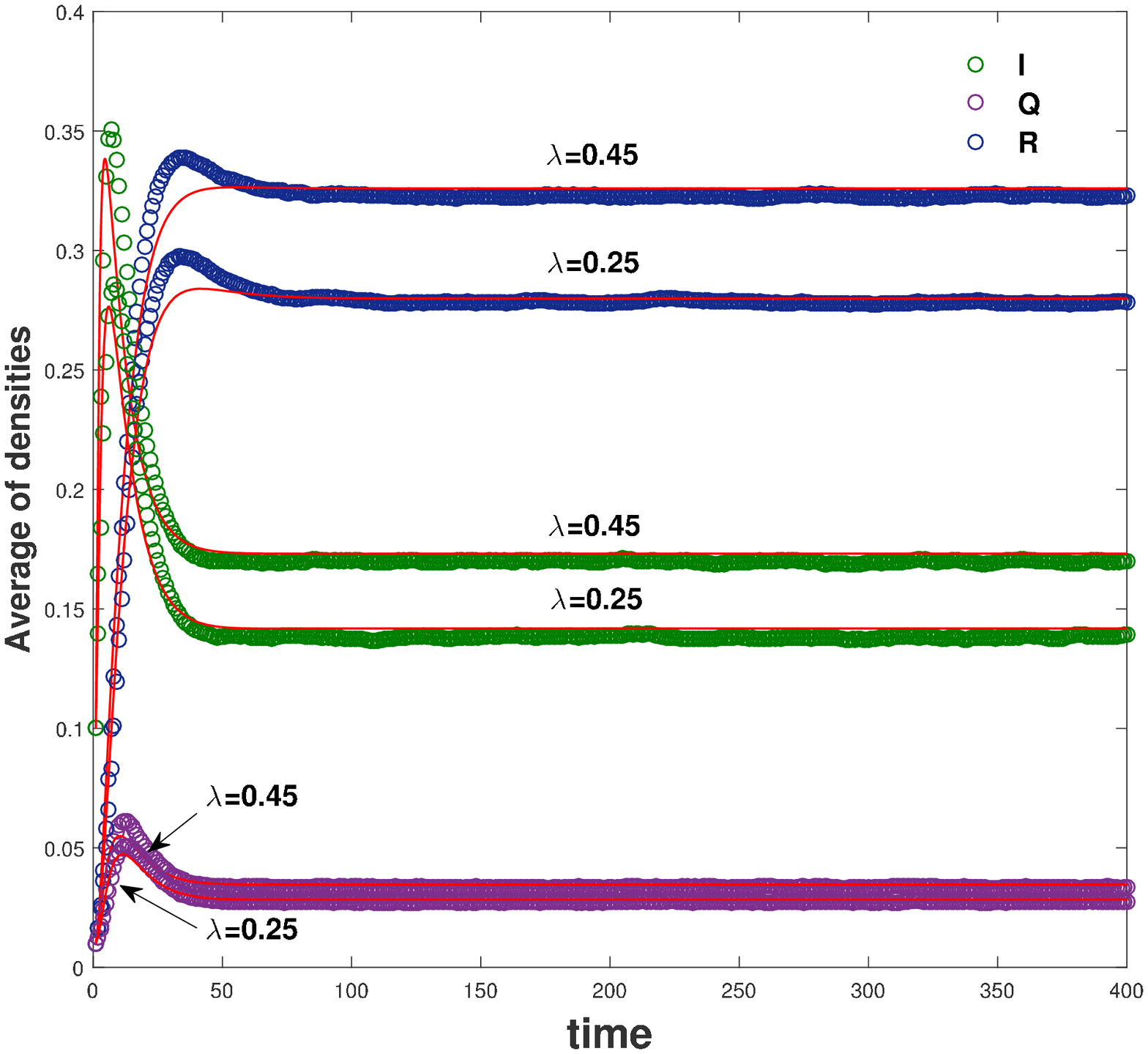}
\label{Fig2(b)}}
~~~~\tiny\subfigure[]{\includegraphics[height=4.5cm,width=7.2cm]{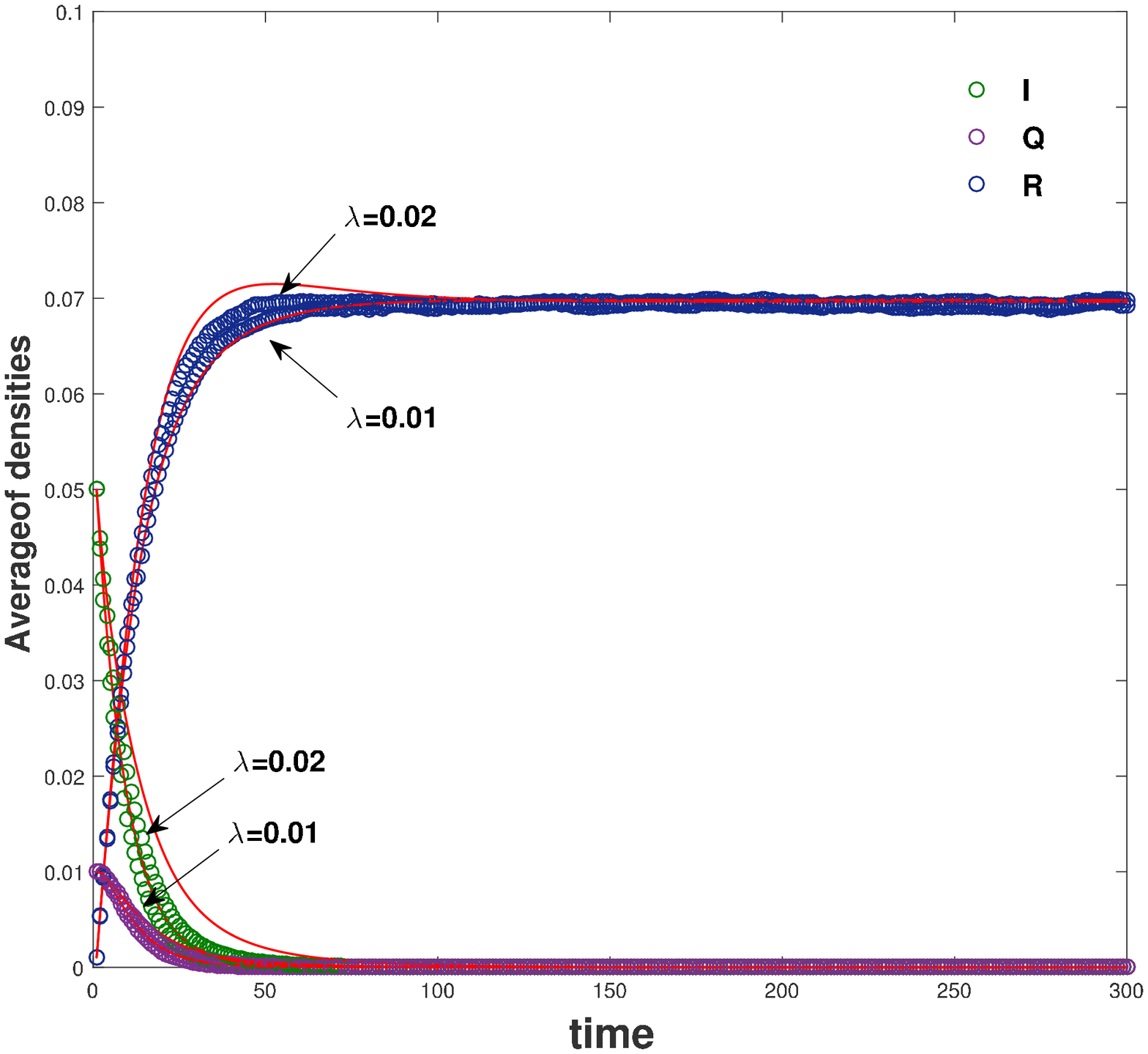}
\label{Fig2(c)}}
~~~~\tiny\subfigure[]{\includegraphics[height=4.5cm,width=7.2cm]{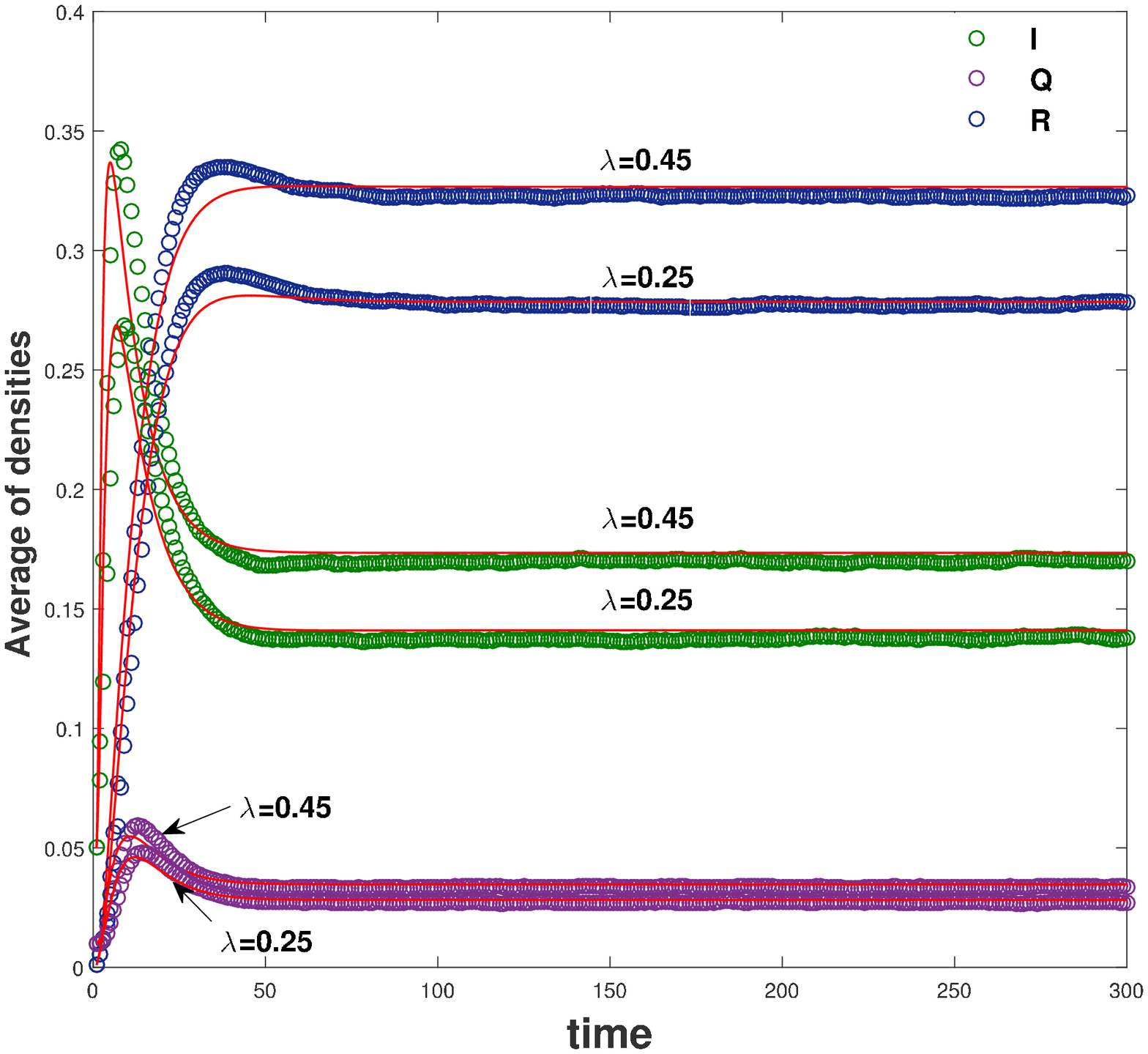}
\label{Fig2(d)}}
\caption{\footnotesize{The comparison of Monte Carlo stochastic simulations (circle) and the mean-field approach (red line) for the SVIQR model ~\eqref{eq33}.
 The parameter values are set as follows:
$\delta $= 0.02, $\alpha$ = 0.01, $\gamma $= 0.05, $\omega $= 0.01, $\mu $= 0.1, $\beta $= 0.05, $b $= 0.2, $d $= 0.05, $\eta = 0.2.$
 \textbf{(a, c)}. The initial fraction of infected and recovered nodes are
set 0.1, 0.01 and 0.05, 0,001, $\lambda$ = 0.01 ($R_{0}$ = 0.4113), $\lambda$ = 0.02 ($R_{0}$ = 0.8225); \textbf{(b, d)}. The initial fraction of infected and recovered nodes are
set 0.1, 0.01 and 0.05, 0,001, $\lambda$ = 0.25 ($R_{0}$ = 10.2814), $\lambda$ = 0.45 ($R_{0}$ = 16.5066). The Monte Carlo stochastic simulations and the mean-field
approach are averaged by 100 realizations. The
agreement between numerical results from the two approaches is very good, which implies that the
analysis based on mean-field approach is very effective.}}
\label{Fig2}
\end{figure}

\subsection{Parameters description and values}

In our simulation, we set the vaccination rate $\mu_{k}=\mu$,
one can set $\mu_{k}$ into different immunization strategies according to the immunization schemes in \cite{35}.
Let $\varphi(k)$ = $k$, $\lambda(k)=\lambda k$.
\begin{figure}[h]
\centering
\tiny \subfigure[]{\includegraphics[height=4.5cm,width=7.2cm]{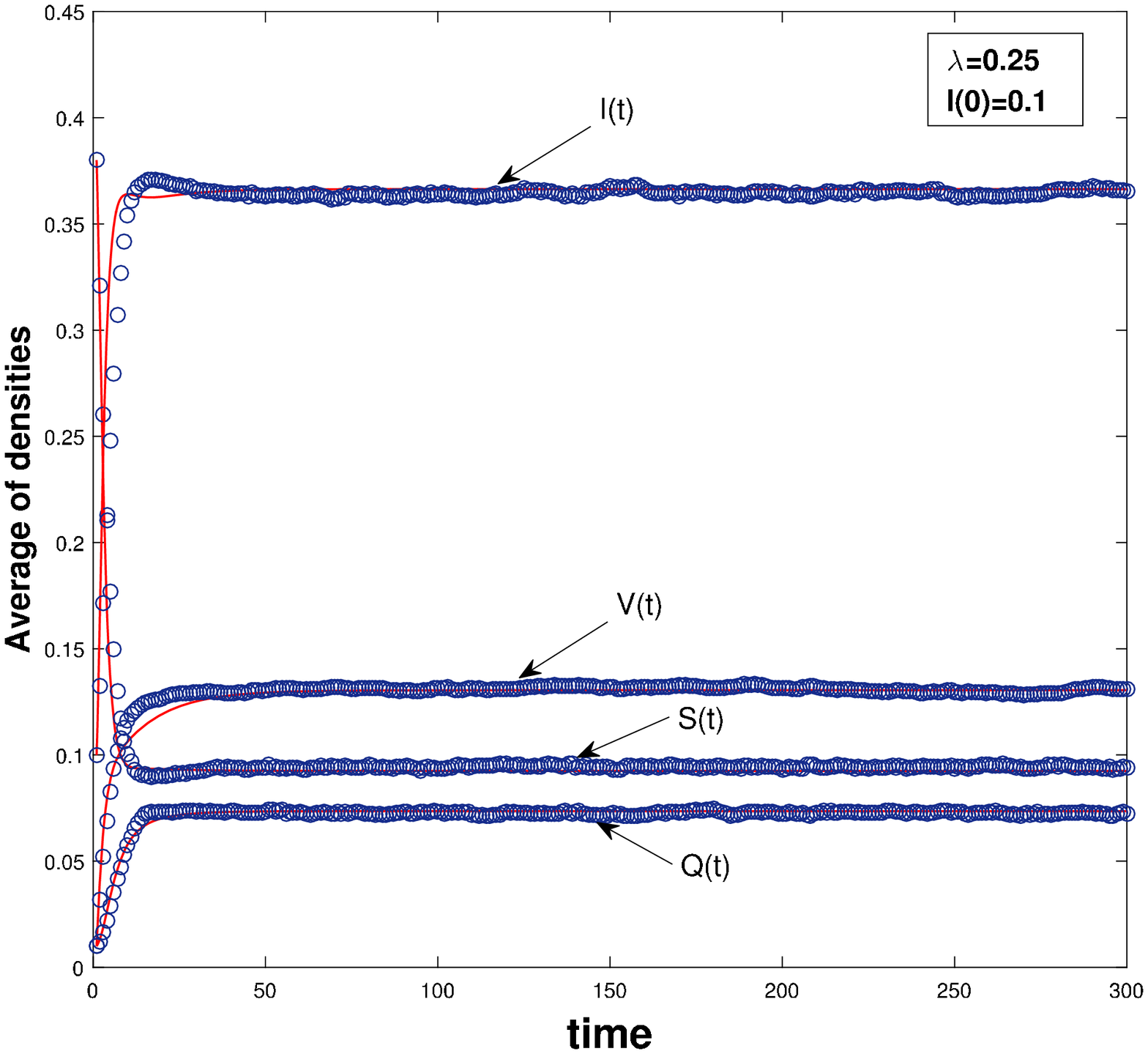}
\label{Fig3(a)}}
~~~~\tiny\subfigure[]{\includegraphics[height=4.5cm,width=7.2cm]{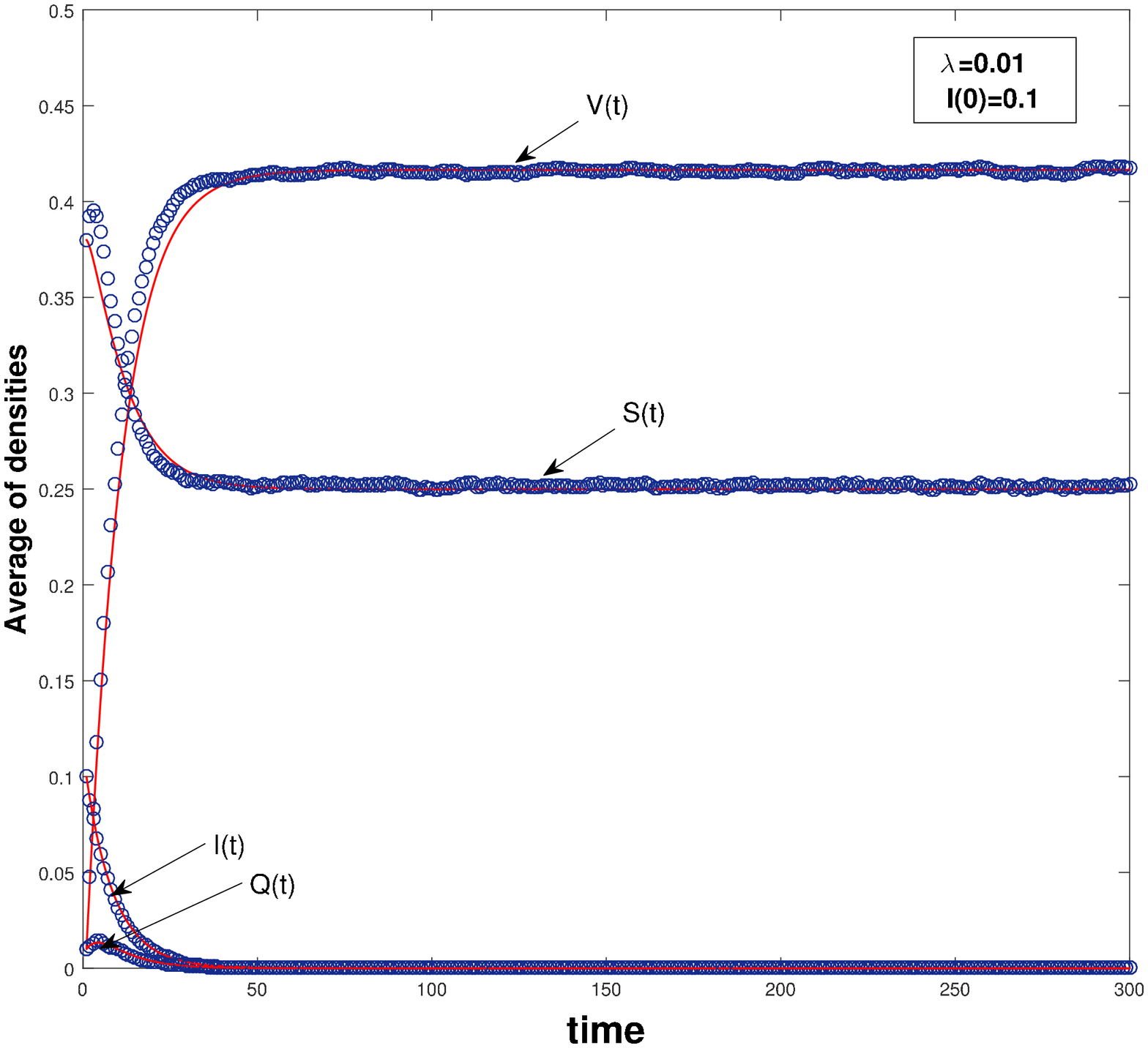}
\label{Fig3(b)}}
~~~~\tiny\subfigure[]{\includegraphics[height=4.5cm,width=7.2cm]{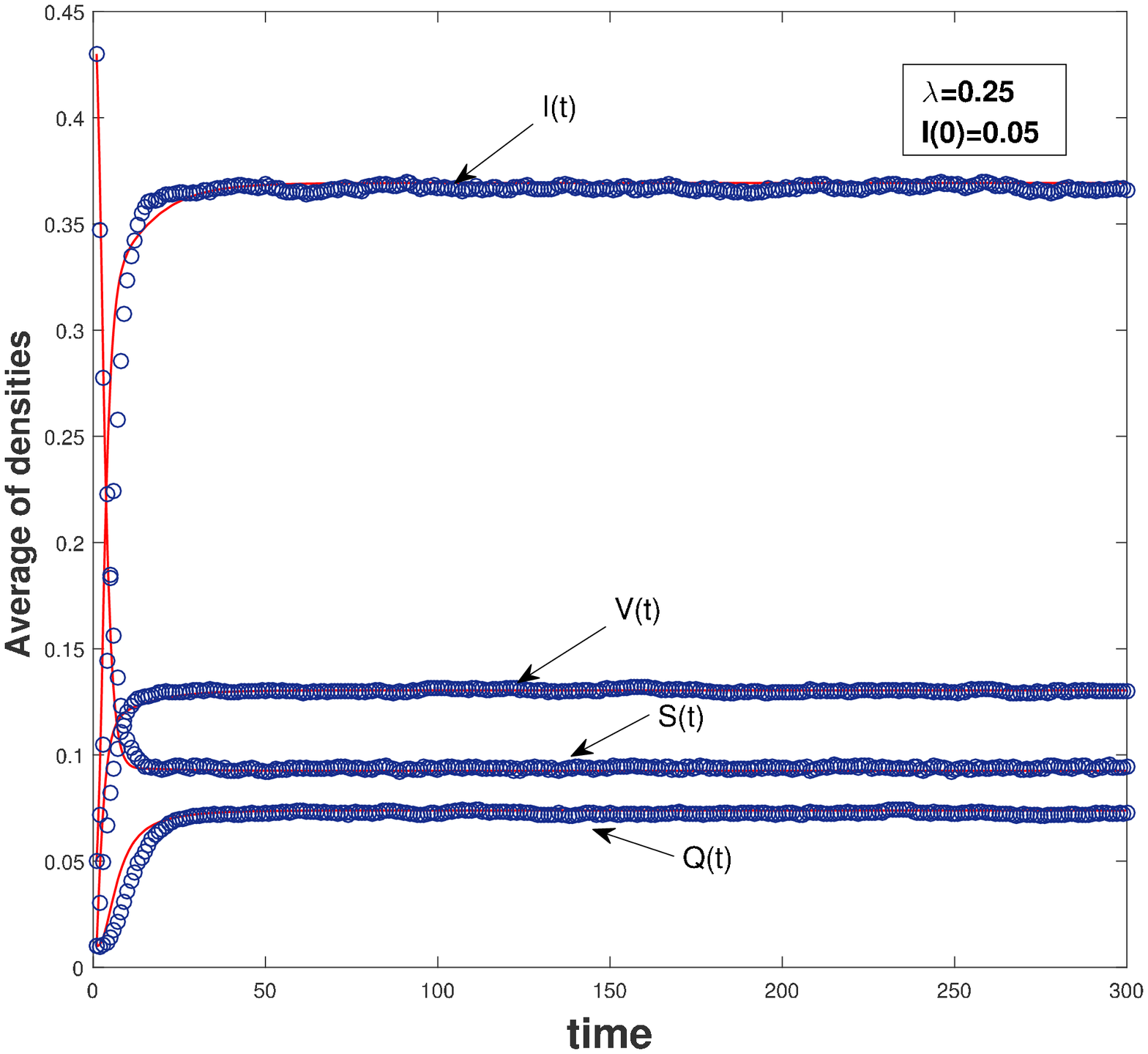}
\label{Fig3(c)}}
~~~~\tiny\subfigure[]{\includegraphics[height=4.5cm,width=7.2cm]{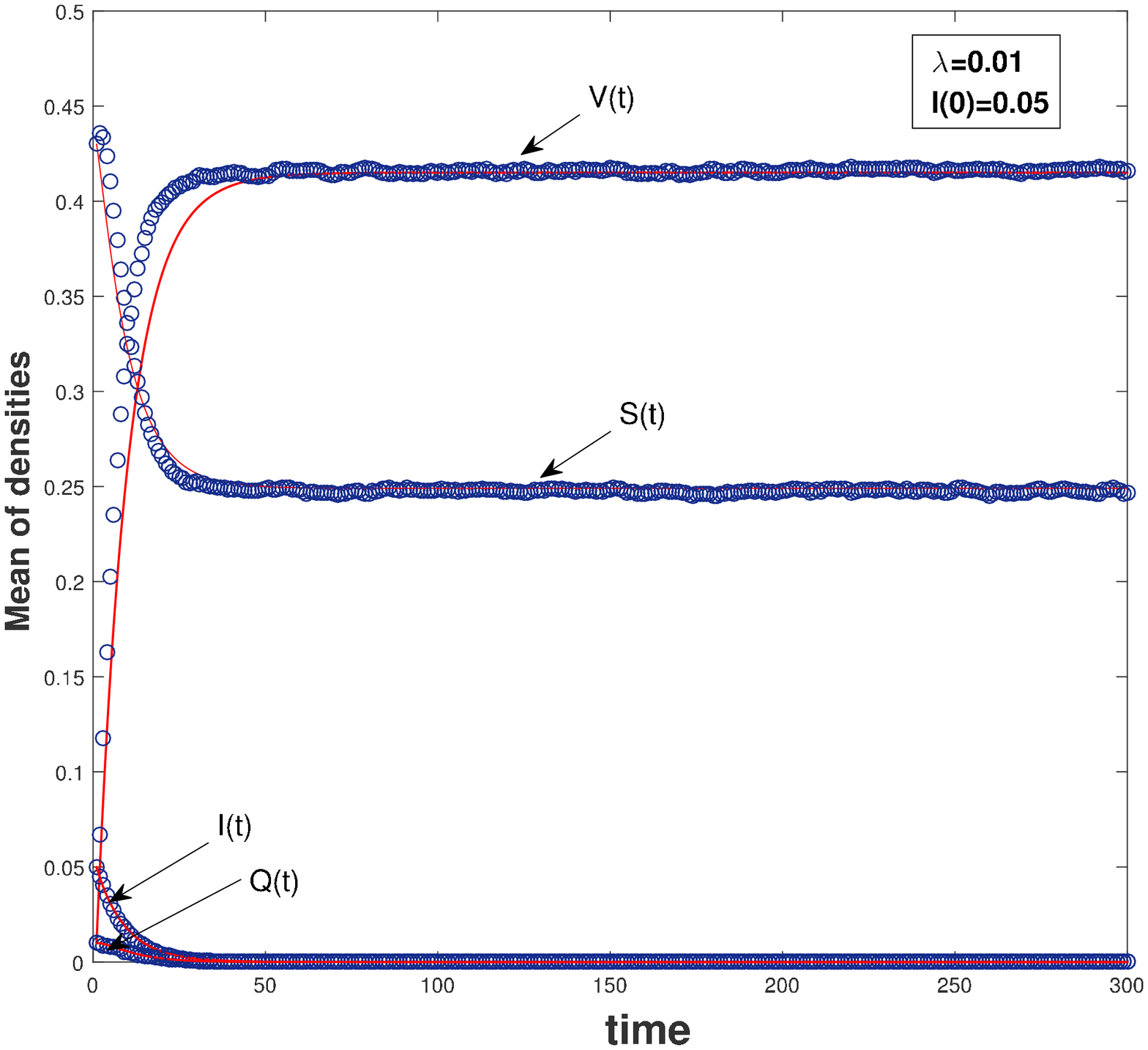}
\label{Fig3(d)}}
\caption{\footnotesize{The comparison of Monte Carlo stochastic simulations (blue circle) and the mean-field approach (red line) for \eqref{eq21}.
 The parameter values are set as follows:
b=0.2, d=0.05, $\mu=0.1$, $\omega=0.01$, $\delta=0.02$, $\gamma= 0.05$, $\eta=0.2$, $\beta=0.05,$ $R^{1}_{0}(\mu_{k})=0.0637<1$.
\textbf{(a, c)}. The initial fraction of infected nodes is
set 0.1 and 0.05, $\lambda$ = 0.25 ($R_{0}$ = 10.5135); \textbf{(b, d)}. The initial fraction of infected nodes is set 0.1 and 0.05,
 $\lambda$ = 0.01 ($R_{0}$ = 0.4205). The Monte Carlo stochastic simulations and the mean-field
approach are averaged by 100 realizations.  The
agreement between numerical results from the two approaches is very good, which implies that the
analysis based on mean-field approach is very effective.}}
\label{Fig7}
\end{figure}
In Fig.~2 and Fig.~3, the initial conditions are $V_{k}(0)=Q_{k}(0)=0.01$ and
$S_{k}(0)=0.5-(V_{k}(0)+I_{k}(0)+Q_{k}(0)+R_{k}(0))$ in system~\eqref{eq33}, $S_{k}(0)=0.5-(V_{k}(0)+I_{k}(0)+Q_{k}(0))$
in system~\eqref{eq21} for any $k=1,2,\cdots,n$. The parameters are set as
$b$=0.2, $d$=0.05, $\delta$=0.02, $\alpha$=0.01, $\gamma$=0.05, $\omega$=0.01, $\mu $=0.1, $\beta$=0.05, $\eta$ = 0.2.

In Fig.~4, Fig.~5, Fig.~6 and Fig.~7, the initial condition are set as $S_{k}(0)=0.3$, $V_{k}(0)=Q_{k}(0)=0.01$.
The parameters are set as $b=0.2$, $d=0.1$, $\alpha=0.01$, $\eta=0.25$.
Other parameters are $\gamma=0.15$, $\lambda$=0.05 in Fig.~4.
And $\omega$=$\delta$=0.01, $\mu$=$\gamma$=0.1, $\beta$=0.15 in Fig.~5 and Fig.~6.
In Fig.~7, we let $\omega$=$\delta$=0.01, $\mu$=0.1, $\beta$=$\gamma$=0.15, $\lambda=0.35$.

In Fig.~7, we study the different quarantine strategies depending on degree $k$ as follows and the average quarantine rate is $\overline{\beta}=\sum_{k}p(k)\beta_{k}$.

\noindent \textbf{Proportional quarantine}: Denoting the quarantine rate $\beta_{k}$ by constant $\beta$;

\noindent \textbf{Targeted quarantine}: By introducing an upper threshold $k_{C}$, such that all nodes with connectivity
$k>k_{C}$ are quarantined, then we defined the quarantine rate by
$\beta_{k}=\left\{\begin{array}{l}
 1, ~k>k_{C}\\
 c,~ k=k_{C}\\
 0,~ k<k_{C}
 \end{array}\right.$, ~where $0 < c \leq 1$.

\noindent \textbf{Acquaintance quarantine}: Choose a random fraction $p$ of the $N$ nodes, the probability that a particular node with
$k$ contacts is selected for quarantine is $\frac{pkp(k)}{\langle k\rangle}=\beta_{k}$.
\begin{figure}[h]
\centering
\tiny\subfigure[]{\includegraphics[height=5cm,width=7.2cm]{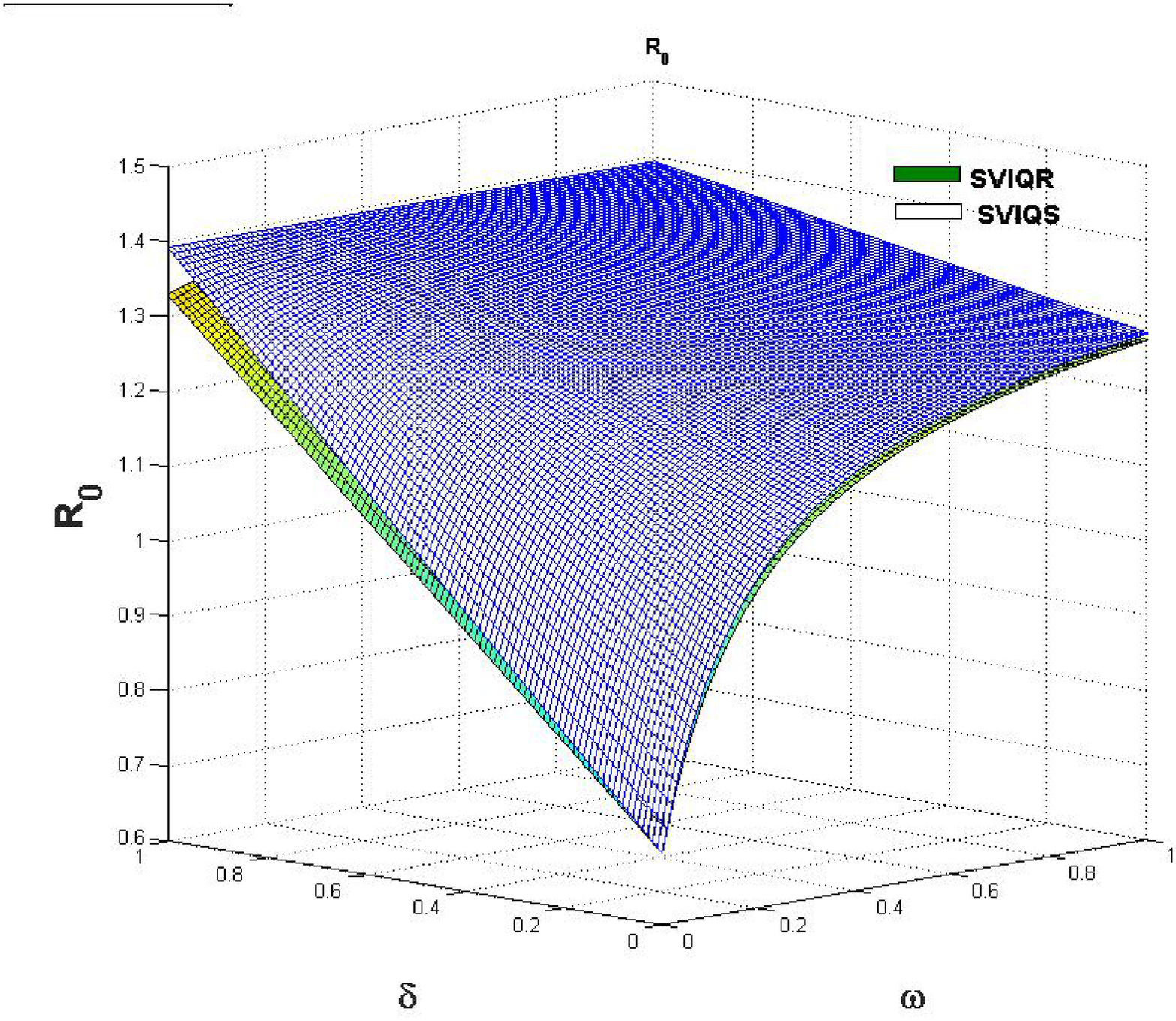}
\label{Fig4(a)}}
~~~~\tiny \subfigure[]{\includegraphics[height=5.1cm,width=7.2cm]{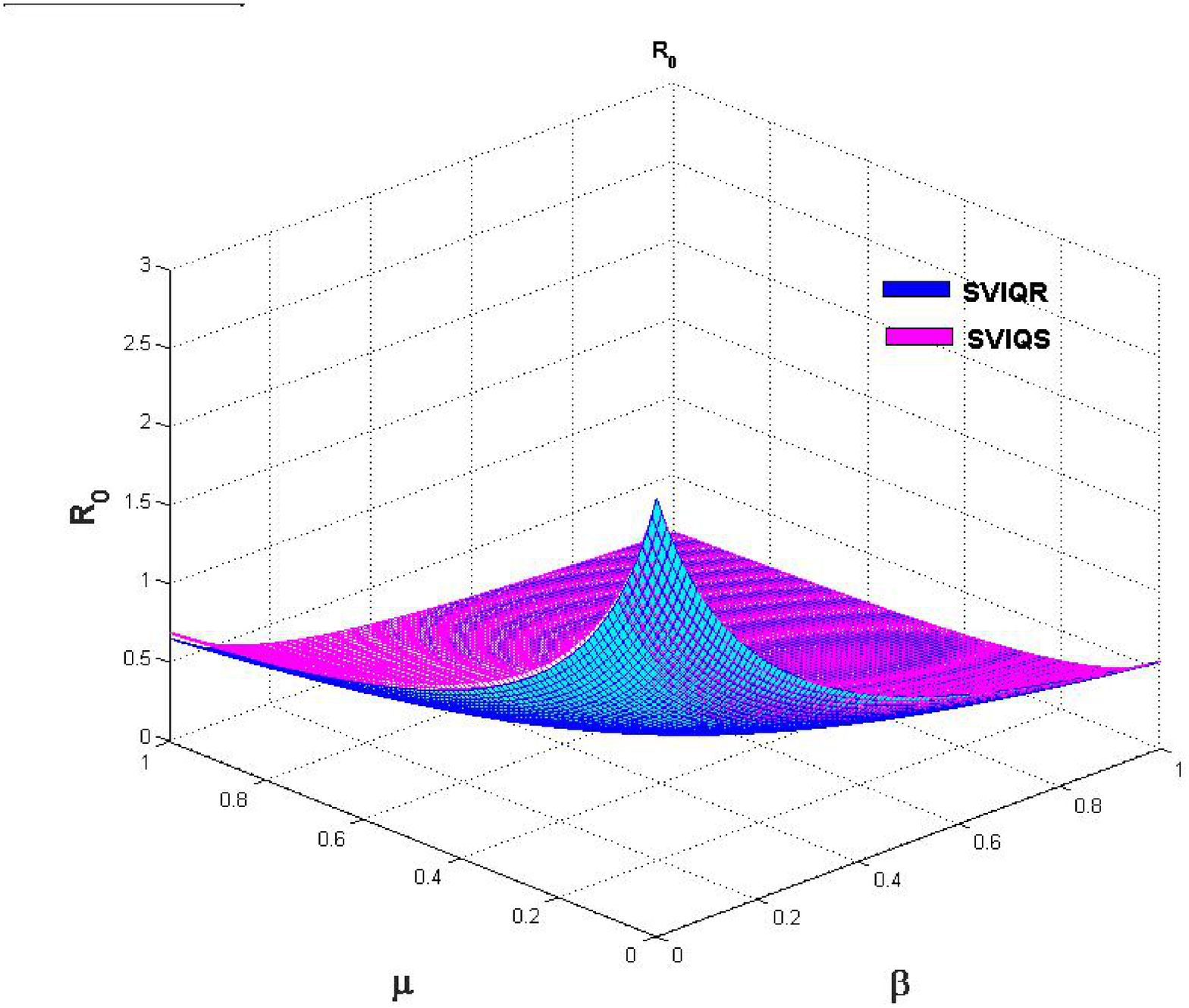}
\label{Fig4(b)}}
\caption{\footnotesize{The combined influence of parameters on $R_{0}$ in SVIQR model~\eqref{eq33} and SVIQS model~\eqref{eq21}.
\textbf{(a)}. The $R_{0}$ in terms of inefficient vaccination rate $\delta$ and relapse rate of vaccinated individuals $\omega$, $\beta=0.1$, $\mu=0.1$ in~\eqref{eq33} and~\eqref{eq21}, respectively;
\textbf{(b)}. The influence of the quarantine rate $\beta$ and vaccination rate $\mu$ on $R_{0}$, $\omega=0.01$, $\delta=0.01$.}}
\label{Fig4}
\end{figure}

\subsection{Numerical results and interpretations}

\subsubsection{Stochastic simulations}

Firstly, we show the comparison of the mean-field approach and Monte Carlo
stochastic simulations for the prediction of the average fraction of infected $I(t)$, quarantined $Q(t)$, recovered individuals $R(t)$
at time $t$ of system~\eqref{eq21} in Fig.~2, and the average fraction of susceptible $S(t)$, vaccinated $V(t)$, infected $I(t)$, quarantined $Q(t)$ individuals of system~\eqref{eq33} in Fig.~3.
 The~\eqref{eq33} and~\eqref{eq21} dynamic process on a
scale-free network proceed with parallel updating. To minimise random fluctuation
caused by the initial conditions, we make average of each state over 100 realizations at each
time step for different initial conditions. From Fig.~2 and Fig.~3, we observe that the
agreement between numerical results from the two approaches is very good, which implies that the
analysis based on mean-field approach is very effective.
Meanwhile, it is clear that the total infected density of~\eqref{eq21} is bigger than the one in~\eqref{eq33}
under the same parameters.
Thus, those diseases which are suitable to~\eqref{eq21} are easier to
spread in population than those diseases which can be depicted by~\eqref{eq33}.
All figures below, our numerical results are mainly obtained from mean-field approach.
\begin{figure}[h]
\centering
\tiny \subfigure[]{\includegraphics[height=4.7cm,width=7cm]{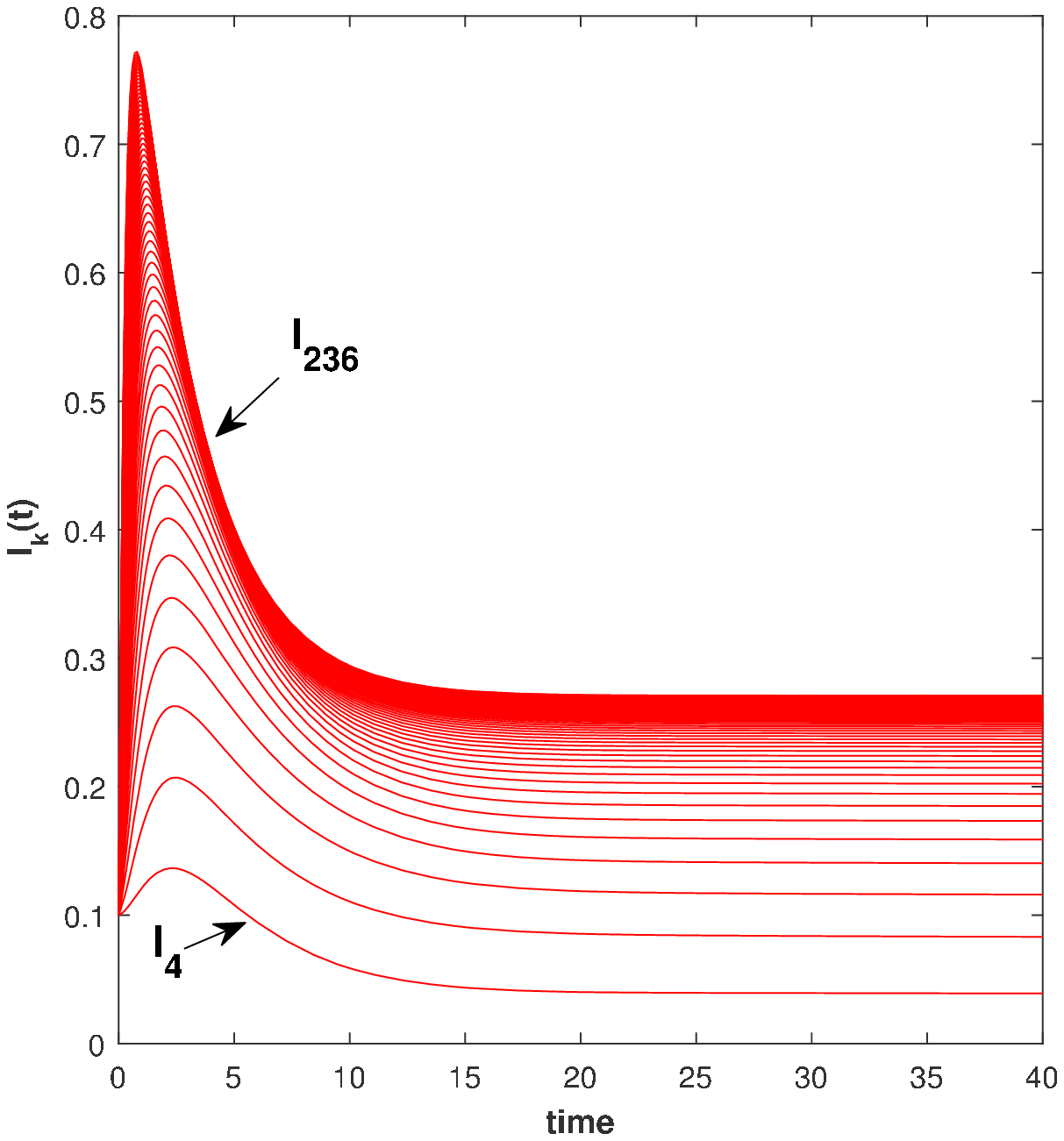}
\label{Fig6(a)}}
~~~~\tiny\subfigure[]{\includegraphics[height=4.7cm,width=7cm]{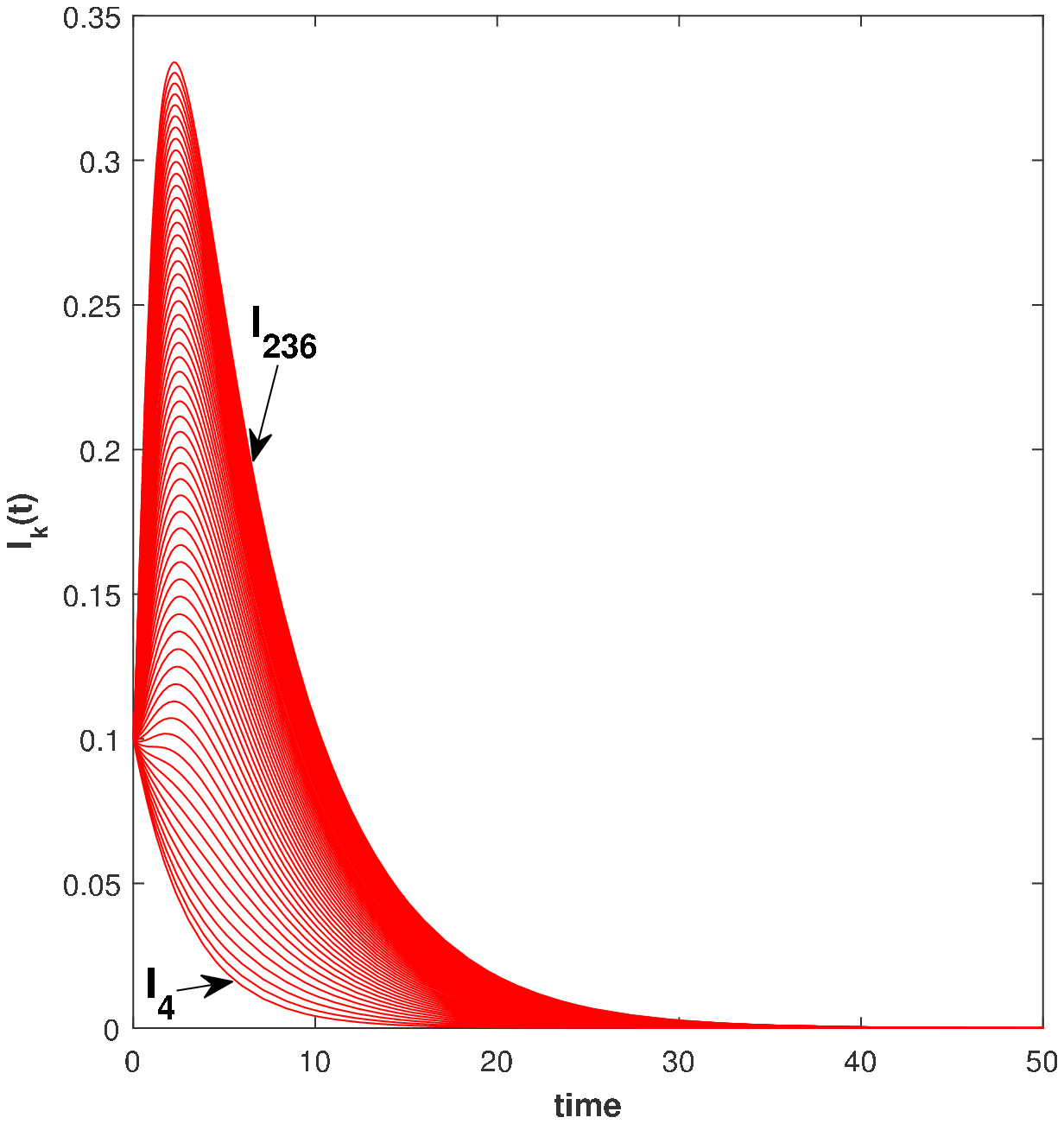}
\label{Fig6(b)}}
~~~~\tiny\subfigure[]{\includegraphics[height=4.7cm,width=7cm]{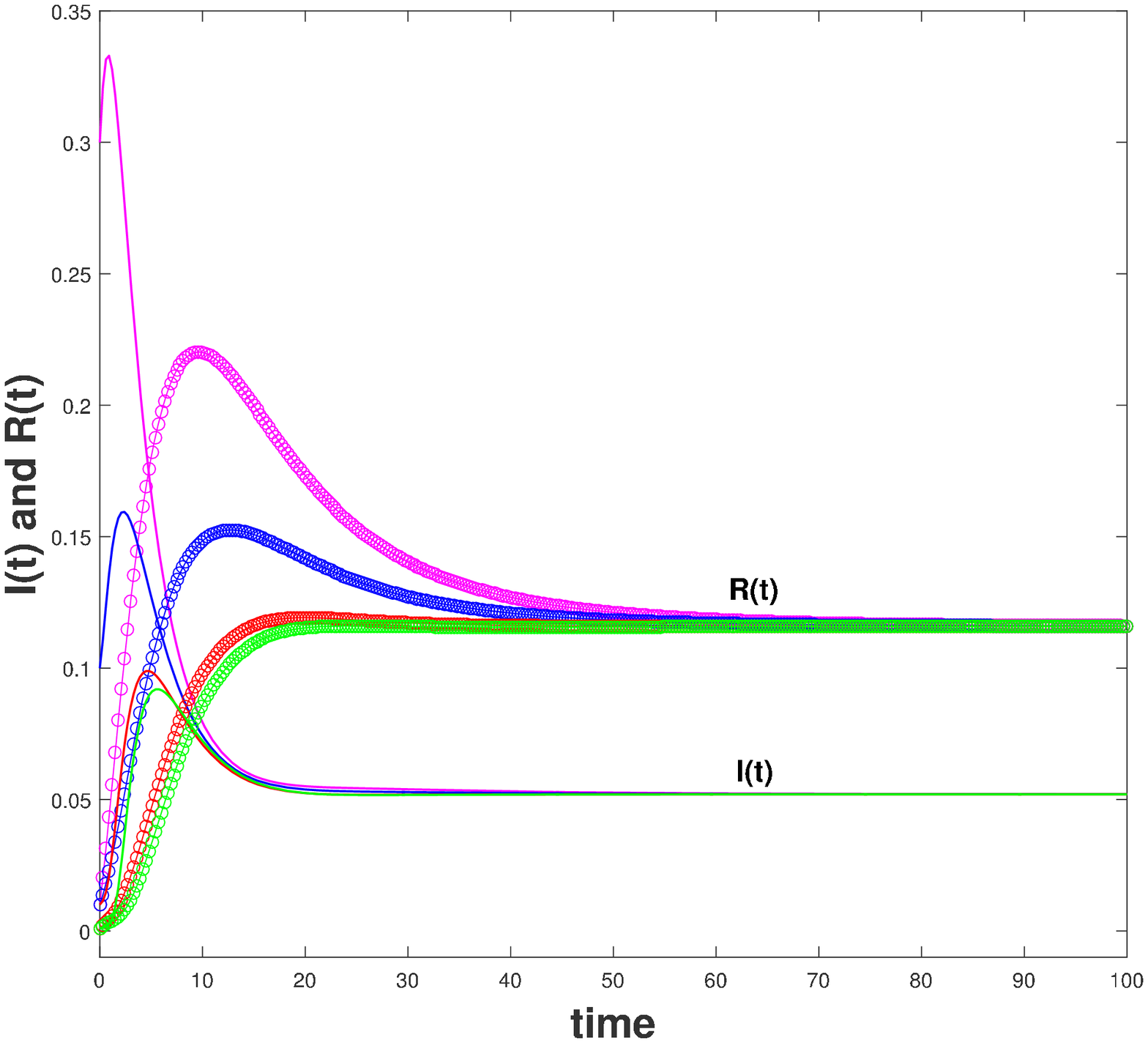}
\label{Fig6(c)}}
~~~~\tiny\subfigure[]{\includegraphics[height=4.7cm,width=7cm]{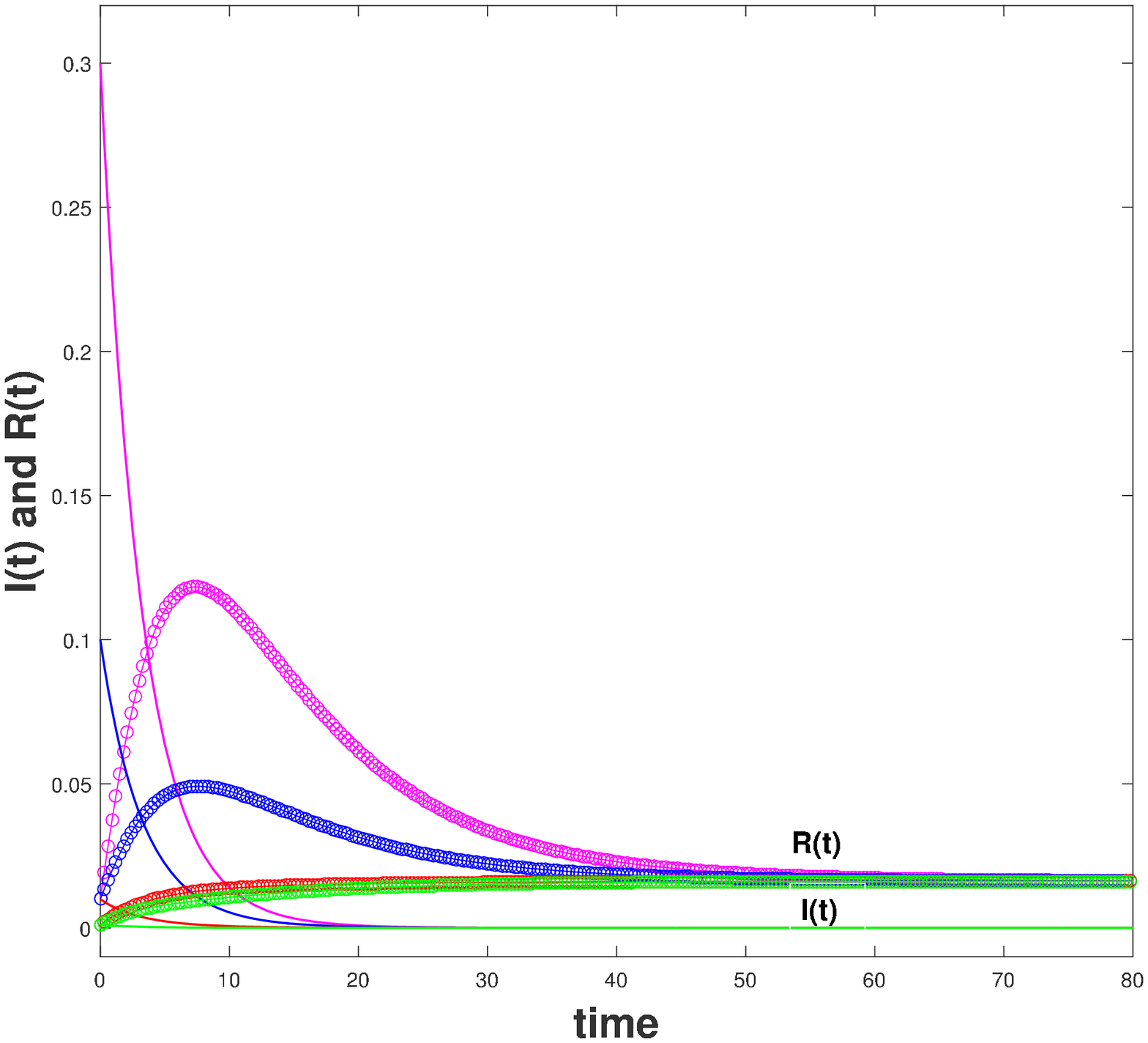}
\label{Fig6(d)}}
\caption{\footnotesize{\textbf{(a, b)}. The densities of infected individuals with different degrees in~\eqref{eq33}.
The lines from bottom to top are $I_{4}(t), I_{8}(t), I_{16}(t),\cdots,I_{236}(t)$, respectively.
\textbf{(c, d)}. The densities of infected (line) and recovered individuals (circle) with different initial conditions in~\eqref{eq33}.
The rose, blue, red and green colors correspond to the initial conditions
 I(0)=0.3 and R(0)=0.01, I(0)=0.1 and R(0)=0.01, I(0)=0.01 and R(0)=0.001, I(0)=0.001 and R(0)=0.001.
\textbf{(a, c)}. $\lambda=0.35$ ($R_{0}=7.3376>1$); \textbf{(b, d)}. $\lambda=0.02$ ($R_{0}=0.4193<1$).}}
\label{Fig7}
\end{figure}

\subsubsection{Mean-field equations simulations}

Fig.~4 illustrates the influence of parameters on $R_{0}$ of SVIQR model \eqref{eq33} and SVIQS model \eqref{eq21}.
All $R_{0}$ increase as the inefficient vaccination $\delta$ and the relapse rate of vaccinated individuals $\omega$ increase.
However, the parameter $\delta$ shows linear positive correction with $R_{0}$, the relationship between $\omega$ and $R_{0}$ is nonlinear.
In addition, the vaccinated rate $\mu$ and quarantine rate $\beta$
have the same effect, their increase will make $R_{0}$ decrease nonlinearly. But, Fig.~4(b)
shows that quarantine plays a more active role than vaccination in controlling the disease.
\begin{figure}[h]
\centering
\tiny \subfigure[]{\includegraphics[height=4.2cm,width=5.1cm]{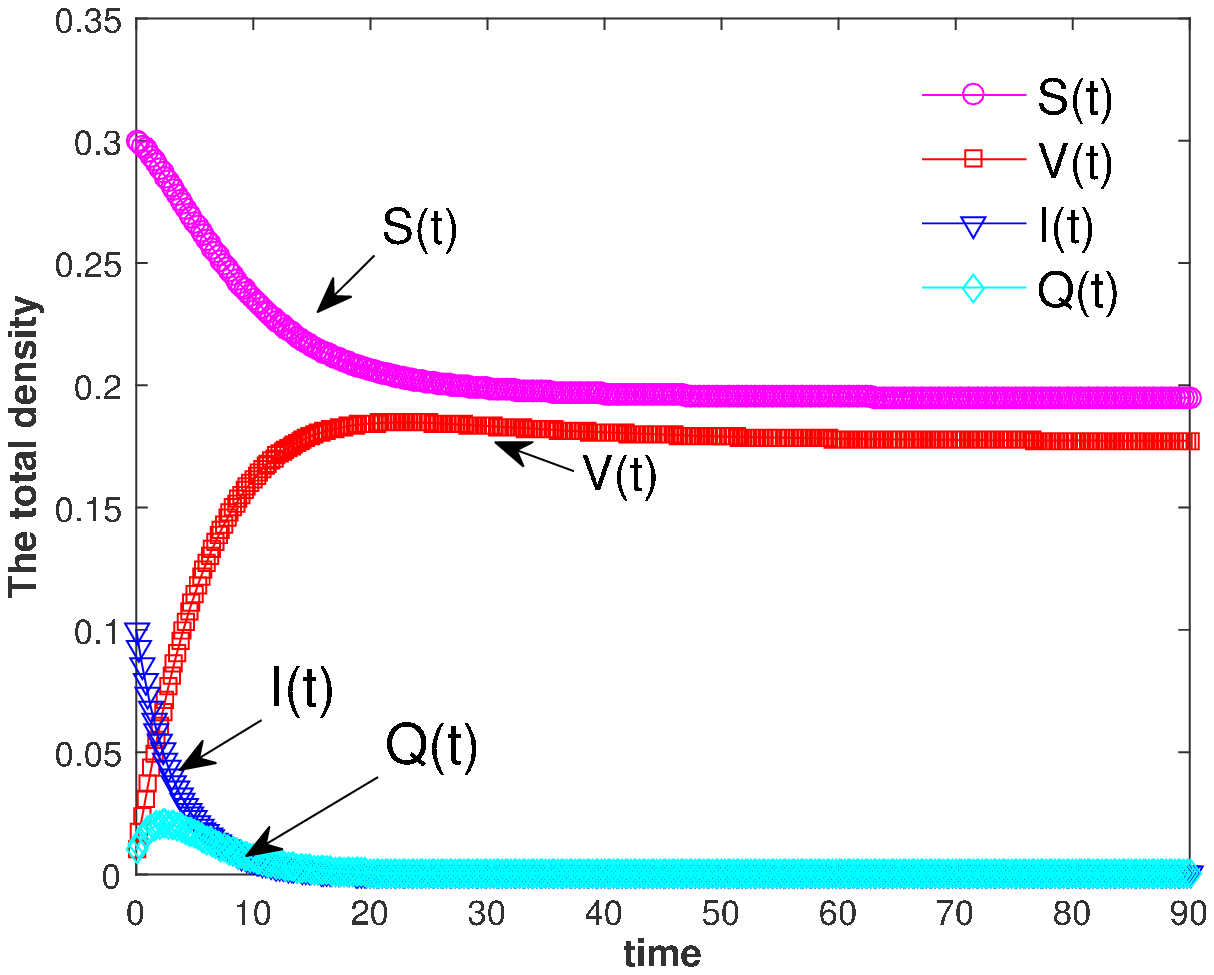}
\label{Fig7(a)}}
~~~~\tiny\subfigure[]{\includegraphics[height=4.2cm,width=5.1cm]{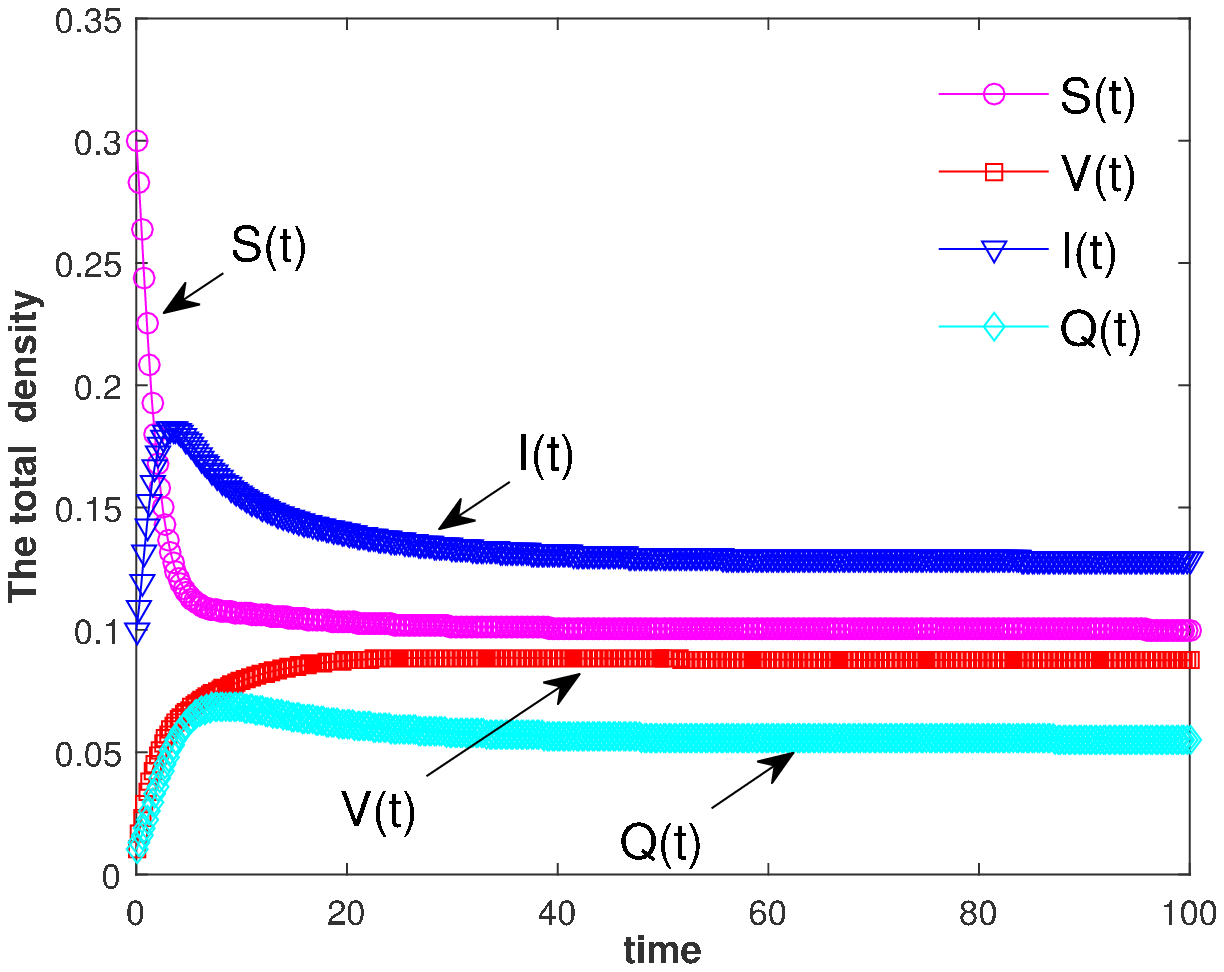}
\label{Fig7(b)}}
~~~~\tiny\subfigure[]{\includegraphics[height=4.2cm,width=5.1cm]{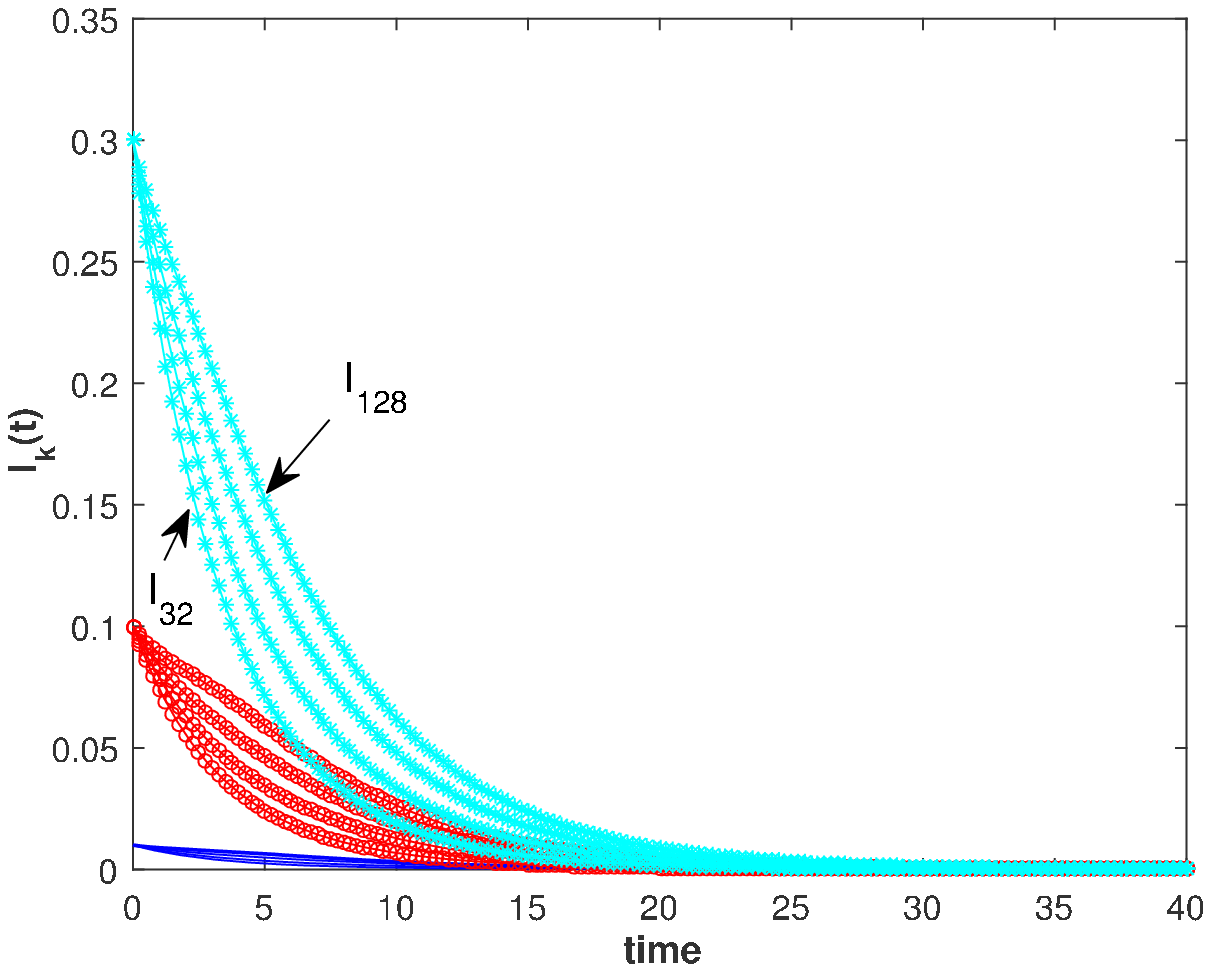}
\label{Fig7(c)}}
~~~~\tiny\subfigure[]{\includegraphics[height=4.2cm,width=5.1cm]{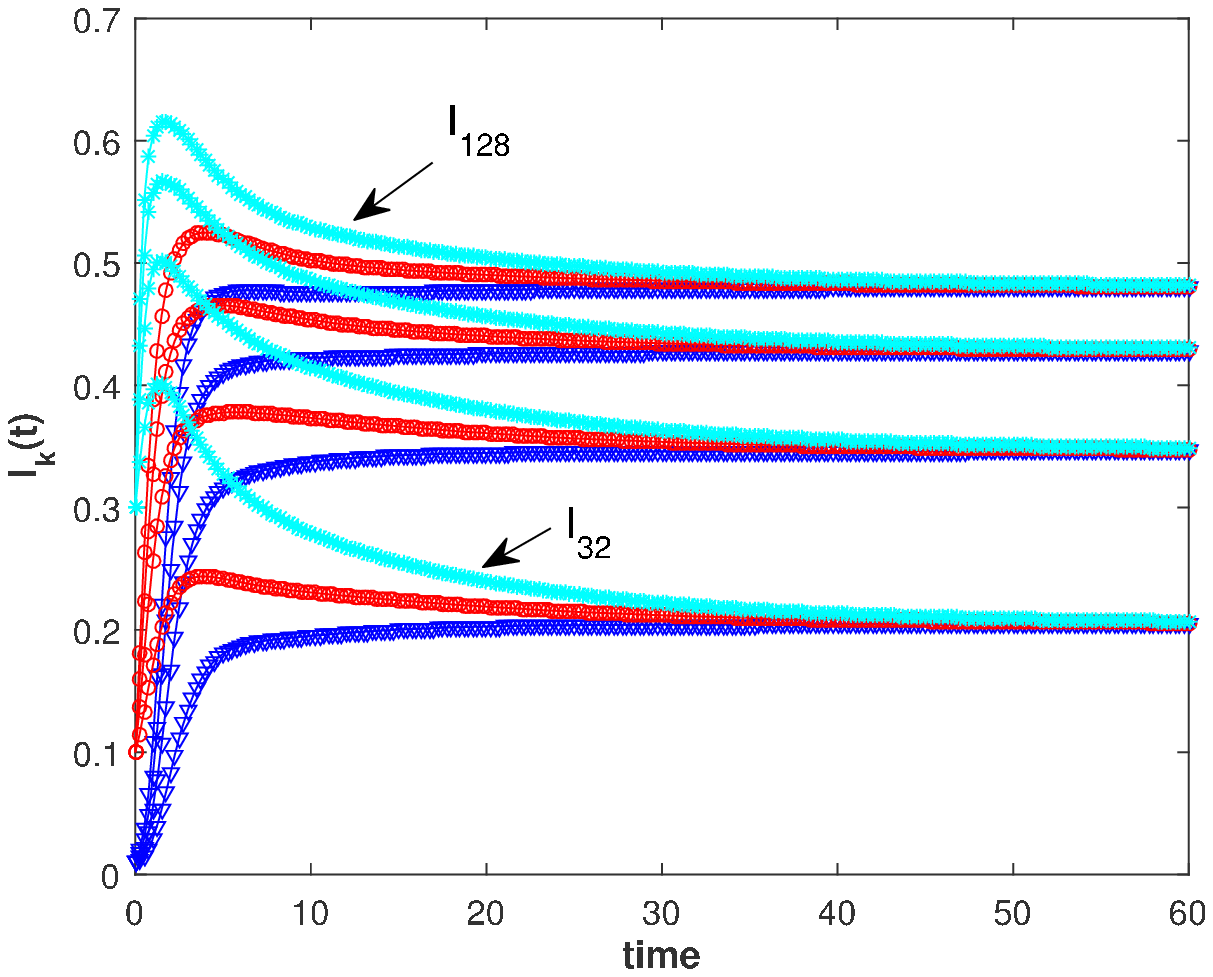}
\label{Fig7(d)}}
~~~~\tiny\subfigure[]{\includegraphics[height=4.2cm,width=5.1cm]{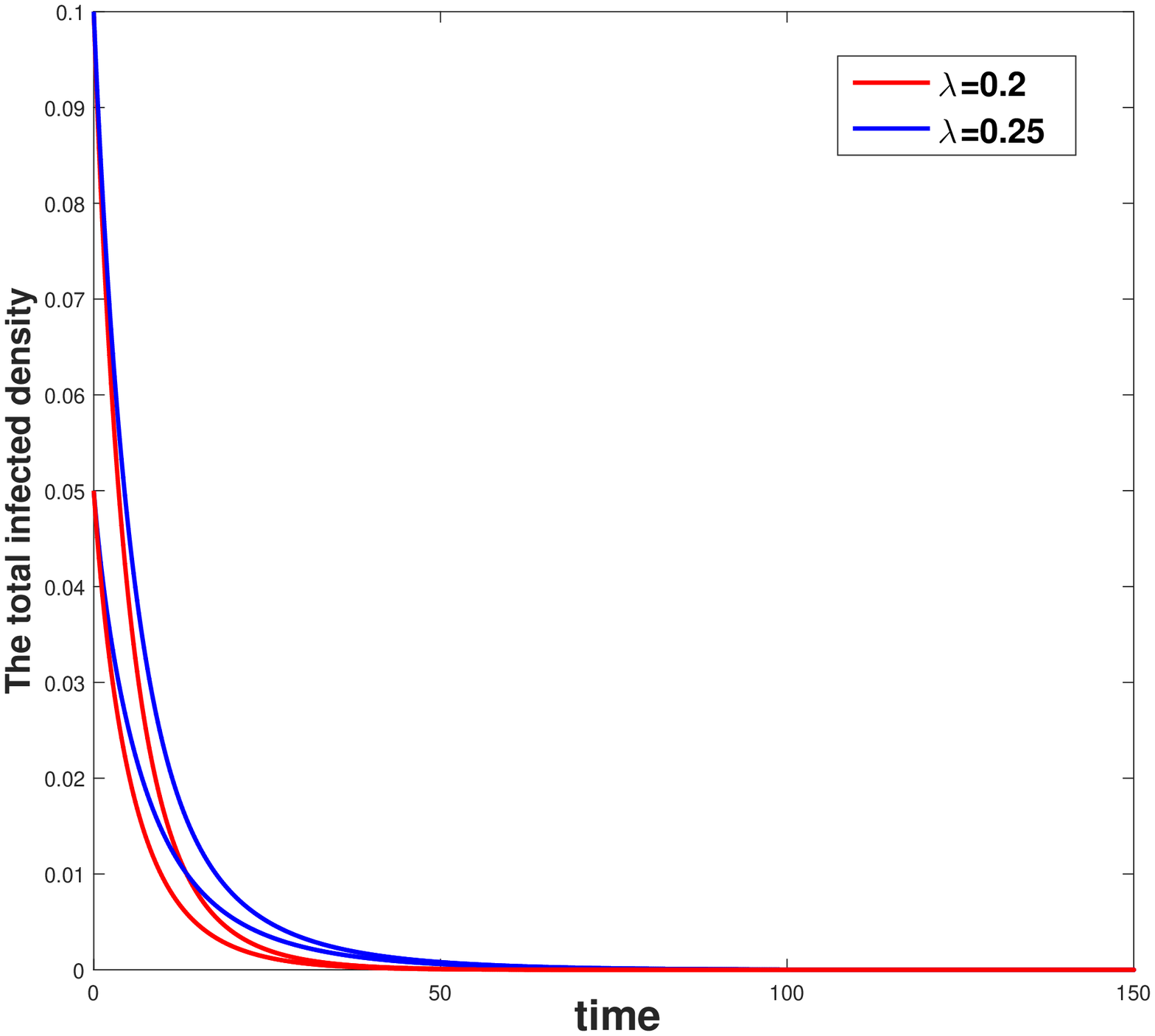}
\label{Fig7(d)}}
~~~~\tiny\subfigure[]{\includegraphics[height=4.2cm,width=5.1cm]{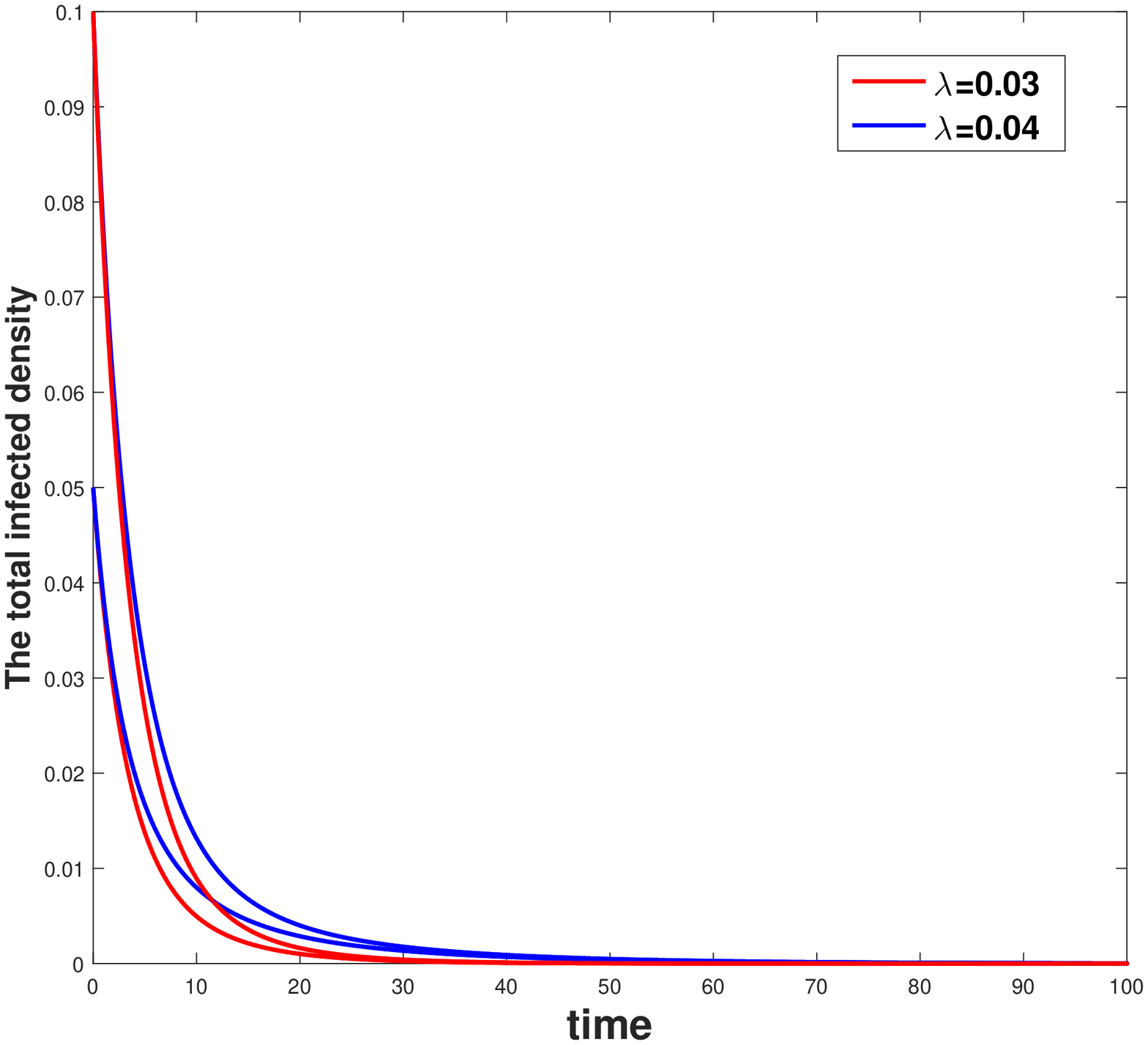}
\label{Fig7(d)}}
\caption{\footnotesize{\textbf {(a, b)}. The total density of each states in SVIQS model~\eqref{eq21} ;
 \textbf {(c, d)}.The influence of initial conditions and degrees on the density of infected individuals in SVIQS model.
The lines with different colors (cyan star, red circle , blue triangle) correspond to I(0)=0.3, 0.1, 0.01 respectively.
Each color line from bottom to top are $I_{32}, I_{64}, I_{96}, I_{128}$.
\textbf{(a, c)} Let $\lambda=0.02$, $R_{0}= 0.4213<1$, $\widetilde{R}_{0}=0.7970$, $R^{1}_{0}(\mu_{k})=0.0197<1$;
\textbf{(b, d)} Let $\lambda=0.35$, $R_{0}=7.3723>1$, $R^{1}_{0}(\mu_{k})=0.0197<1$.
\textbf{(e)} Let $\varphi(k)=k$, for red line, $R^{1}_{0}(\mu_{k})=0.0197$,
 $\lambda=0.03$, $R_{0}=0.6320$, $\widetilde{R}_{0}=1.1956$,
 for blue line $\lambda=0.04$, $R_{0}=0.8427$, $\widetilde{R}_{0}=1.5942$;
 \textbf{(f)} Let $\varphi(k)=k^{0.5}$, $R^{1}_{0}(\mu_{k})=0.0197$, for blue line,
 $\lambda=0.25$, $R_{0}=0.8189$, $\widetilde{R}_{0}=1.5493$,
 for the red line $\lambda=0.2$, $R_{0}=0.6551$, $\widetilde{R}_{0}=1.2394.$}}
\label{Fig7}
\end{figure}
\begin{figure}[h]
\centering
\tiny \subfigure[]{\includegraphics[height=5cm,width=7cm]{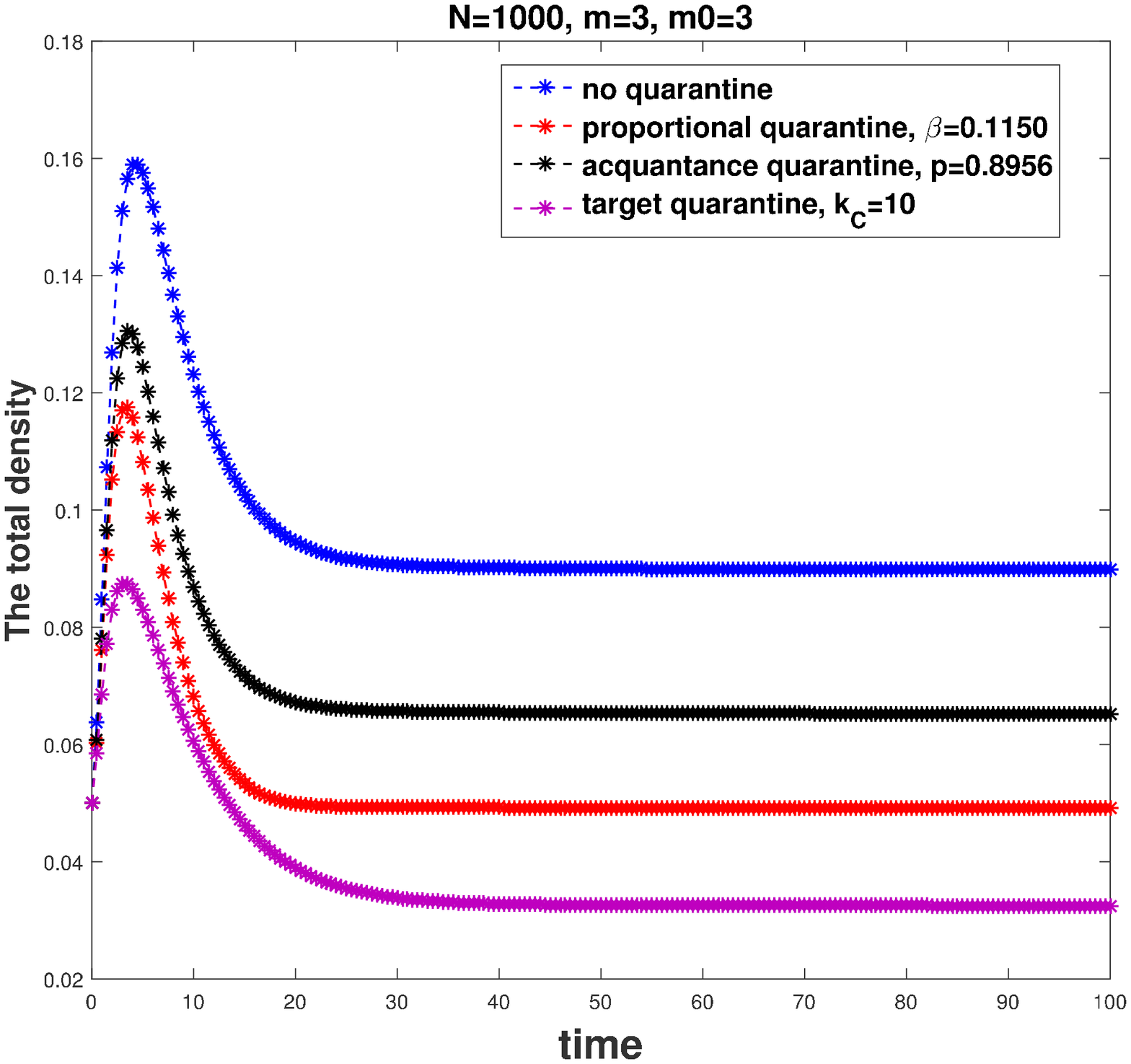}
\label{Fig4(a)}}
~~~~\tiny\subfigure[]{\includegraphics[height=5cm,width=7cm]{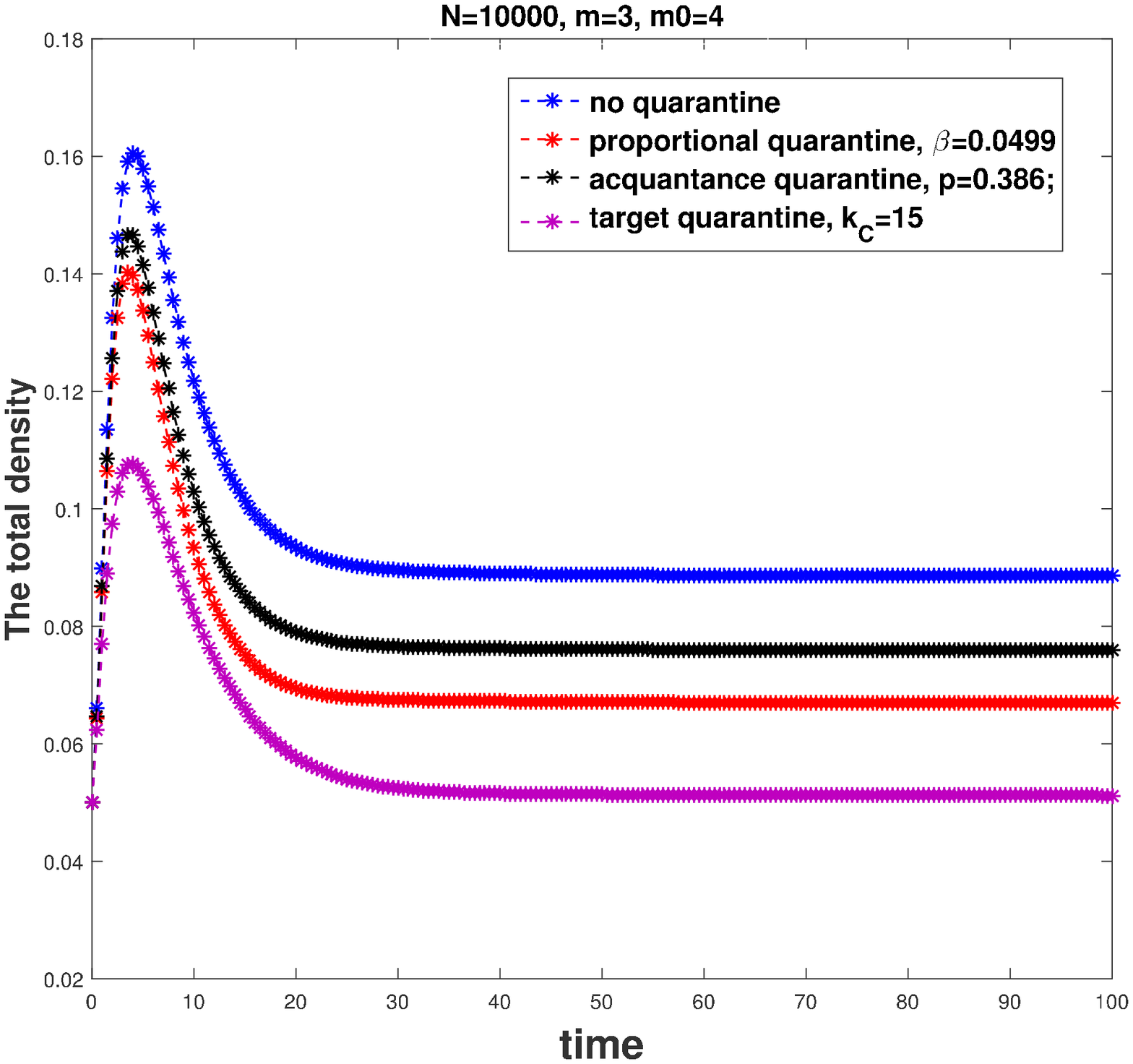}
\label{Fig4(b)}}
\caption{\footnotesize{Comparison of the effectiveness of different quarantine schemes about total infected infection in different networks.
\textbf{(a)} For target quarantine, $k_{C}=10$; for acquaintance quarantine, $l=0.8956$; for proportional quarantine, $\beta=0.1150$.
\textbf{(b)} For target quarantine, $k_{C}=15$; for acquaintance quarantine, $l=0.386$; for proportional quarantine, $\beta=0.0499$.
 All of them can make sure that the average quarantine rate is equal to proportional quarantine $\beta$.}}
\label{Fig4}
\end{figure}

We compute numerical simulations to test and complement the main results from Theorem 4.4 to Theorem 4.10 by Fig.~5 and Fig.~6.
In addition, we study the influence of degree and initial conditions on epidemic dynamics.

For system~\eqref{eq33}, we let $\lambda=0.02$ in Fig.~5(b),(d) to make sure
$R_{0}<1$ and let $\lambda=0.35$ in Fig.~5(a),(c) ensure $R_{0}>1$,
they depict that the disease-free and endemic equilibria of ~\eqref{eq33} are globally asymptotically stable, respectively,
which correspond to the results of Theorem 4.4 and Theorem 4.5.

 For system~\eqref{eq21}, we set the parameters to ensure $R^{1}_{0}(\mu_{k})\leq1$.
In Fig.~6(a),(c), we let $\lambda=0.02$ such that $R_{0}<1$,
it is clear that the disease-free equilibrium is globally asymptotically stable although the condition $\omega\geq\eta=\gamma$ is not satisfied.
Let $\lambda=0.35$ in Fig.~6(b),(d), we get $R_{0}>1$ in system~\eqref{eq21},
the disease reaches at endemic equilibrium point and does not disappear even if $\omega\geq\eta=\gamma$ is invalid.
Obviously, Fig.~6(a)-(d) show the supplementary results of Theorem 4.7, Theorem 4.8 and Theorem 4.10, namely,
we obtain that even if the conditions $\omega\geq\eta=\gamma$ is not satisfied, the
disease-free and endemic equilibria are still globally asymptotically stable as long as the condition $R^{1}_{0}(\mu_{k})\leq1$ is satisfied.
Furthermore, Fig.~6(e), (f) show that under the condition of $R^{1}_{0}(\mu_{k})\leq1$, the disease equilibrium of system~\eqref{eq21}
is global asymptotical stability for $R_{0}<1$, although $\widetilde{R}_{0}>1$.

In addition, from Fig.~5(c),(d) and Fig.~6(c),(d), we can observe that the initial conditions have almost no influence
on the stationary fraction of infected individuals. If $R_{0}<1$ (and also satisfies $R^{1}_{0}(\mu_{k})\leq1$ in SVIQS model), no matter
how many infected individuals initially exist, the disease eventually dies out quickly.
If $R_{0}>1$ (and also satisfies $R^{1}_{0}(\mu_{k})\leq1$ in SVIQS model), the disease persists on a unique positive state.
Fig.~5(a),(b) and Fig.~6(c),(d) depict that the larger degree $k$, the higher steady levels $I_{k}$ with larger degree $k$.

There are little research on quarantine strategies depending on degree $k$,
we now investigate the quarantine strategies about quarantine rate $\beta_{k}$ in SVIQR model~\eqref{eq33} in Fig.~7(a),(b).
The proportional quarantine, the targeted quarantine ($k_{C}$ =10 and $k_{C}$ =15) and the acquaintance quarantine ($p$= 0.8956, $p$= 0.386) have been defined in Section 5.1.
From Fig.~7, we compare the total infected density $I(t)$ in \eqref{eq33} among no quarantine and the other three quarantine schemes in the same network.
 It shows that all three quarantine schemes are more effective than the case without quarantine. In particular,
the targeted quarantine scheme is the most effective than other schemes
 for the same average quarantine rate $\overline{\beta}=\beta=0.1150$ in Fig.~7(a) and $\overline{\beta}=\beta=0.0499$ in Fig.~7(b).


\section{Conclusions}

We highlight in this article the importance of imperfect vaccination and quarantine in the control of the propagation of different diseases.
We rely on a general network-based SVIQR model~\eqref{eq33} that characterizes the infectious diseases which can lead to permanent natural immunity.
And, we also develop and analyze a general network-based SVIQS epidemic model~\eqref{eq21} to study those disease that can not lead to permanent immunity to infection.
Our two general models combined with demographics, general degree-related imperfect vaccination, quarantine, as well as general infectivity.

We obtain expression for $R_{0}$, the key parameter related to epidemic dynamics, which governs the asymptotic behavior of model.
The $R_{0}$ in system~\eqref{eq21} has the same expression with the $R_{0}$ in system~\eqref{eq33}
 when $\alpha$ is zero.
We have showed that these two $R_{0}$ are closely related to the topology of networks and some parameters by theoretical analysis and simulations.
 In particularly, the effects of vaccination and quarantine on epidemic dynamics have been discussed.
It seems that the quarantine rate $\beta_{k}$ and the vaccination rate $\mu_{k}$ have the same effects,
 because their increase will make $R_{0}$ decrease. However, we find that quarantine plays a more active role
than vaccination in controlling disease.
In addition, it seems that the inefficient vaccination rate to infection $\delta$ and relapse rate of vaccinated individuals $\omega$ also have the same effect,
while the parameter $\delta$ shows linear positive correction with $R_{0}$, the influence of $\omega$ on $R_{0}$ presents nonlinear positive correction.

For SVIQR model~\eqref{eq33}, both analytical and numerical results emphasize that the basic reproduction number $R_{0}$ shows the sharp threshold property
complete governing the global dynamics of model, although the imperfect vaccination and quarantine are considered in scale-free networks.
Therefore, there does not exists endemic equilibrium for $R_{0}<1$, namely, this system cannot undergo backward bifurcation.
More specifically, by constructing Lyapunov function, we prove that the disease-free and endemic equilibria are globally asymptotically stable.
Obviously, the global stability of disease-free equilibrium for $R_{0}<1$
also gives an alternative approach to exclude the occurrence of backward bifurcation.
Hence, our results improve and extend the results of \cite{26} and \cite{33}.

For SVIQS model~\eqref{eq21}, we have performed a qualitative
analysis to show that~\eqref{eq21} provides interesting dynamical behaviors.
In the presence of multiple endemic
equilibria, $R_{0}<1$ might not be sufficient to eliminate the disease.
We have derived the condition $\min\{R^{1}_{0}(\mu_{k}),R^{2}_{0}(\mu_{k})\}>1$ under which system~\eqref{eq21} may exist multiple endemic
equilibria (it may happen in the backward bifurcation scenario) for $R_{0}<1$.
This condition is not only dependent on parameters but also network structure.
As showed in Theorem 4.2, under condition $R^{1}_{0}(\mu_{k})\leq 1$, there is no
endemic equilibrium for $R_{0}<1$,
the system exhibits forward bifurcation for $R^{1}_{0}(\mu_{k})\leq 1$.
We make use of some lemmas to show that the disease-free equilibrium is globally asymptotically stable if $\widetilde{R}_{0}<1$ without any assumption on parameters.
However, this result can not exclude the stability of endemic equilibria which may exist for
$R_{0}<1$. In order to investigate the global stability of equilibria,
we assume that $\omega\geq\eta=\gamma$ satisfying $R^{1}_{0}(\mu_{k})\leq1$ and show the disease-free equilibrium is globally asymptotically stable if $R_{0}<1$.
Meanwhile, by applying a monotone iterative technique, it is shown that if $R_{0}>1$, the
epidemic equilibrium $E^{*}$ is globally attractive under assumption $\omega\geq\eta=\gamma$.
In addition, as is observed from the simulations, we have shown that
the disease-free equilibrium is global asymptotically stable for $R_{0}<1$, and
the endemic equilibrium is also globally asymptotically stable
for $R_{0}>1$ as long as the condition $R^{1}_{0}(\mu_{k})\leq1$ is satisfied, although the condition $\omega\geq\eta=\gamma$ may not be satisfied.

We also consider the different quarantine strategies under ignoring the quarantine
time to infected individuals in SVIQR model~\eqref{eq33} in different networks.
The result of simulation shows that quarantine is important in control the disease,
in particularly, the target quarantine is the most effective in control the disease.

Numerical simulations have confirmed and complemented the theoretical results.
 This paper provides a concise mathematical proof of the global dynamics,
and this approach can be generalized to other epidemic models on heterogeneous networks.
It would be interesting to further consider more dynamical behavior of SVIQS model~\eqref{eq21} without the assumption of parameters
and the optimal vaccination and quarantine strategies on complex networks.
We hope to tackle these questions in the future.

\section*{Acknowledgments}
\indent
This work was jointly supported by the NSFC under grants 11572181 and 11331009.
SC was also supported with funding from ARC Linkage grant LP130101055.
SC is also grateful to Drs. Debora Correa and Jack Moore at Univ.
of Western Australia for their kind help and encouragement.

\small{}
\end{document}